\documentclass[letterpaper,11pt]{article}
\title{Polyhedral Clinching Auctions for Indivisible Goods}

\author{Hiroshi Hirai \thanks{Nagoya University, Nagoya, Japan. hirai.hiroshi@math.nagoya-u.ac.jp} \and Ryosuke Sato \thanks{Keio University, Yokohama, Japan. ryosuke.sato.517@gmail.com}}

\usepackage[top=1truein,bottom=1truein,left=1truein,right=1truein]{geometry}
\usepackage{amsmath}
\usepackage{amsfonts}
\usepackage{amssymb}
\usepackage{amsthm}
\usepackage{algorithm}
\usepackage{algorithmic}
\usepackage{amsthm}
\usepackage{comment}
\usepackage[hidelinks]{hyperref}
\usepackage{tikz}
\usetikzlibrary{arrows.meta,positioning}
\usetikzlibrary{calc}

\begin{document}
\theoremstyle{definition}
\newtheorem{definition}{definition}[section]
\newtheorem{proposition}[definition]{Proposition}
\newtheorem{lemma}[definition]{Lemma}
\newtheorem*{main}{Main Theorem}
\newtheorem*{theorem2}{Theorem}
\newtheorem{theorem}[definition]{Theorem}
\newtheorem{corollary}[definition]{Corollary}
\newtheorem{remark}[definition]{Remark}
\newtheorem*{rem}{Remark}
\newtheorem{fact}[definition]{Fact}
\newtheorem{claim}[definition]{Claim}
\newtheorem{observation}[definition]{Observation}
\newtheorem{assmp}[definition]{Assumption}
\newtheorem{example}[definition]{Example}

\maketitle
\begin{abstract}
In this study, we propose the polyhedral clinching auction for indivisible goods, 
which has so far been studied for divisible goods. As in the divisible setting by Goel et al. (2015), 
our mechanism enjoys incentive compatibility, individual rationality, and Pareto optimality, 
and works with polymatroidal environments. A notable feature for the indivisible setting 
is that the whole procedure can be conducted in time polynomial of the number of buyers and goods. 
Moreover, we show additional efficiency guarantees, recently established by Sato for the divisible setting: 
The liquid welfare (LW) of our mechanism achieves more than 1/2 of the optimal LW, and that the social welfare is more than the optimal LW. 
\end{abstract}

\section{Introduction}
The theoretical foundation for budget-constrained auctions is an unavoidable step toward further social implementation of auction theory. A representative example of such auctions is ad auctions (e.g., Edelman et al. \cite{EOS2007}), where advertisers naturally have budgets for their advertising costs. However, it is well known  \cite{DLN2012,DL2014}  that designing auctions with budget constraints 
is theoretically difficult: Any budget-feasible mechanism cannot achieve the desirable goal of satisfying all of incentive compatibility (IC), individual rationality (IR), and constant approximation to the optimal social welfare (SW).

When budgets are public, 
Dobzinski et al. \cite{DLN2012} proposed a budget-feasible mechanism that 
builds on the clinching framework of Ausubel \cite{A2004}. 
Their mechanism satisfies IC, IR, and Pareto Optimality (PO), 
a weaker notion of efficiency than SW. 
They also showed that their mechanism is the only budget-feasible mechanism satisfying IC, IR, and PO. 
These results have inspired subsequent research \cite{BHLS2015, DHS2015, FLSS2011, GMP2014, GMP2015, GMP2020, HS2022} for extending their mechanism to various settings.

{\it Polyhedral clinching auction} by Goel et al. \cite{GMP2015} is the most outstanding of these, 
and can even be applied to complex environments expressed by polymatroids. 
Their mechanism is a clever fusion of auction theory and polymatroid theory. 
This has brought  further extensions, such as concave budget constraints \cite{GMP2014} and two-sided markets \cite{HS2022, S2023}. 
Particularly, a recent result by Sato \cite{S2023} established a new type of efficiency guarantees in the mechanism. 
Thus, the polyhedral clinching auction is a standard framework for the theory of budget-constrained auctions. 

These results of the polyhedral clinching auctions are all restricted to auctions of divisible goods, though many auctions deal with indivisible goods.
In this study, we address the polyhedral clinching auction for indivisible goods to enlarge its power of applicability.

\subsubsection*{Our Contributions}

We propose the polyhedral clinching auction for indivisible goods, 
based on the framework of the one for divisible goods in Goel et al. \cite{GMP2015}. 
Our mechanism exhibits the desirable properties expected from theirs.
That is, it satisfies IC, IR, and PO and works with polymatroidal environments. 
This means that it is applicable to a wide range of auctions, such as 
multi-unit auctions in Dobzinski et al. \cite{DLN2012}, matching markets in Fiat et al. \cite{FLSS2011}, 
ad slot auctions in Colini-Baldeschi et al. \cite{BHLS2015}, and video-on-demand in Bikhchandani et al. \cite{BSV2011}.
In addition, a promising future research is two-sided extensions of our results, 
as already proceeded for the divisible settings in Hirai and Sato \cite{HS2022} and Sato \cite{S2023}.
Particularly, such an extension includes the reservation exchange markets in Goel et al. \cite{GLMNP2016}\ ---\ a setting of two-sided markets for display advertising. 

As in the divisible setting in \cite{GMP2015}, each iteration of our mechanism can be done in polynomial time. 
A notable feature specific to the indivisible setting is iteration complexity. The total number of the iterations is also polynomially bounded in the number of buyers and the goods. 
Thus, the whole procedure can be implemented in polynomial time. 

In addition to the above PO, we establish two types of efficiency guarantees.
The first one is that 
our mechanism achieves {\it liquid welfare} (LW) more than 1/2 of the optimal LW.
This is the first LW guarantee for clinching auctions with indivisible goods. Here LW \cite{DL2014, ST2013} is a payment-based efficiency measure for budget-constrained auctions, and is defined as the sum of the total admissibility-to-pay, which is the minimum of valuation of the allocated goods (willingness-to-pay) and budget (ability-to-pay). 
Our LW guarantee is understood as an indivisible and one-sided version of the one recently established by Sato \cite{S2023} for the divisible setting. The notable point is that these LW guarantees 
hold for such general auctions, even in indivisible setting, while other existing work \cite{DL2014} on 
LW guarantees for clinching auctions is only limited to simple settings of multi-unit auctions. 

The second one is that our mechanism achieves SW more than the optimal LW.
This type of efficiency guarantee, which compares SW with the optimal LW, 
was initiated by Syrgkanis and Tardos \cite{ST2013} in the Bayesian setting, and was recently obtained by Sato \cite{S2023} for clinching auctions with divisible goods in the prior-free setting.
In budget-constrained auctions, modifying the valuations to make 
the market non-budgeted is often considered; see, e.g., Lehmann et al. \cite{LLN2006}. 
If each buyer's valuation is modified to the budget-additive valuation, 
then LW is interpreted as SW. In this approach, the optimal LW is used as the target value of SW 
and can be thought as a reasonable benchmark of SW. 
Thus, this guarantee provides another evidence for high efficiency of our mechanism.

\subsubsection*{Our Techniques}
{\it Tight sets lemma} \cite{GMP2014,GMP2015} characterizes the dropping of buyers and the 
final allocation, and has been a powerful tool for efficiency guarantees of  
polyhedral clinching auctions for divisible goods. 
For showing efficiency guarantees (PO, LW, SW) mentioned above, 
we establish a new and the first tight sets lemma for indivisible setting. 
Although our (hard-budget) setting is a natural indivisible version of 
the one in Goel et al. \cite{GMP2015},  
the indivisibility causes complications in various places 
and prevents straightforward generalization in both its formulation and proof. 
We utilize the notions of {\it dropping prices} and {\it unsaturation}
by Goel et al. \cite{GMP2014} invented for a more complex divisible setting 
(concave-budget setting), and formalize and prove our indivisible tight sets lemma.

Even with our new tight sets lemma,
the indivisibility still prevents a straightforward adaptation
of previous techniques showing efficiency guarantees,
especially the LW guarantee.
The proof of the above 2-approximation LW guarantee
is based on the idea of Sato \cite{S2023} for divisible setting, 
and is obtained by establishing the inequality
\[
{\rm LW}^{\rm M} \geq p^{\rm f}(N) \geq {\rm LW}^{\rm OPT} - {\rm LW}^{\rm M},
\]
where ${\rm LW}^{\rm M}$ and $p^{\rm f}(N)$ 
are the LW value and the total payments, respectively, in our mechanism, 
and ${\rm LW}^{\rm OPT}$ is the optimal LW value.
In the divisible setting of \cite{S2023}, the second inequality is obtained
by using the LW-optimal allocation to provide a lower bound on future payments.
However, due to indivisibility, this approach does not fit in our setting.
Instead, we introduce a new technique of lower bounding the future payments
via virtual buyers and the associated virtual optimal LW allocation
to overcome the difficulty.
This new technique is interesting in its own right,
and expected to be applied to LW guarantees of other auctions.

\subsubsection*{Other Related Works}

Auction of indivisible goods is ubiquitous in the real-world. For unbudgeted settings, its theory already has a wealth of knowledge; see, e.g., Krishna \cite{K2010} and Nisan et al.~\cite{NRTV2007}. 
Auction with (poly)matroid constraints was initiated by Bikhchandani et al. \cite{BSV2011}. They considered buyers who have concave valuations and no budget limits. Our framework captures a budgeted extension of their framework if all buyers have additive valuations within their budgets. 
Candogan and Peke\v{c} \cite{CP2018} developed a network-flow approach 
to implement the VCG outcome when buyers have 2-feature valuations, generalizing the concave valuations in Ausubel \cite{A2004}. 
Our result is also an extension of \cite{A2004}, but is based on (poly)matroid theory and focuses on constraints rather than valuations.
For budgeted auctions, Dobzinski et al.~\cite{DLN2012} proposed the adaptive clinching auction and showed that it has IC, IR, and PO. Later, Fiat et al.~\cite{FLSS2011} and Colini-Baldeschi et al~\cite{BHLS2015} extended their mechanism to a market represented by a bipartite graph. Our framework is also viewed as a generalization of theirs to polymatroidal~settings.

LW was introduced independently and simultaneously by Dobzinski and Leme \cite{DL2014} and Syrgkanis and Tardos \cite{ST2013}. The existing LW guarantees for auctions (with public budgets) are as follows: For clinching auctions, Dobzinski and Leme \cite{DL2014} showed that the adaptive clinching auction in Dobzinski et al. \cite{DLN2012} achieves 2-approximation to the optimal LW. Recently, Sato \cite{S2023} showed that the polyhedral clinching auction in Hirai and Sato \cite{HS2022} achieves 2-approximation to the optimal LW even under polymatroidal constraints. Our results are viewed as an indivisible and one-sided version of his results. For unit price auctions, Dobzinski and Leme \cite{DL2014} also showed that their unit price auction achieves 2-approximation to the optimal LW. Later, Lu and Xiao \cite{LX2015} proposed another unit price auction and improved the guarantee to $(1+\sqrt{5})/2$. 
It is an interesting research direction to incorporate polymatroidal constraints with their mechanisms, for which our results may help.

\subsubsection*{Organization of this paper}
The rest of the paper is organized as follows. In Section~2, we introduce our model. In Section~3, we propose our mechanism and provide some basic properties. In Section~4, we analyze the structure of our mechanism and obtain the tight sets lemma. In Section~5, we provide the efficiency guarantees for our mechanism with respect to PO, LW, and SW. In Section~6, we conclude this paper and present some future directions. In Appendix, we give a further discussion of our tight sets lemma and provide omitted proofs.

\subsubsection*{Notation}
Let $\mathbb R$, $\mathbb R_+$, $\mathbb R_{++}$
denote the set of real numbers, nonnegative real numbers, positive real numbers, respectively, and let $\mathbb Z_{+}$ denote the set of nonnegative integers.
For a set $N$, let $\mathbb R^N$ (or $\mathbb R^N_+$, $\mathbb Z^N_+$) denote the set of 
all functions from $N$ to $\mathbb R$ (or $\mathbb R_+$, $\mathbb Z_+$). 
For $x\in \mathbb R^N$ (or $\mathbb R^N_+$, $\mathbb Z^N_+$), 
we often denote $x(i)$ by $x_i$, and write it as $x= (x_i)_{i\in N}$. 
For $S\subseteq N$, let $x(S)$ denote the sum of $x(i)$ over $i\in S$, i.e., $x(S) := \sum_{i\in S} x(i)$.
In addition, for $S \subseteq N$, let $x|_{S}$ denote the restriction of $x$ to $S$.
We often denote a singleton $\{i\}$ by $i$ and a set $\{1,2,\ldots,k\}$ by $[k]$.

\section{Our Model}
Consider a market with multiple buyers and one seller who plays the role of the auctioneer. 
The seller auctions multiple units of a single indivisible good. 
Let $N:=\{1,2,\ldots,n\}$ be the set of buyers. 
Each buyer $i\in N$ has three positive real numbers $v_i, v'_i, B_i \in \mathbb R_{++}$, where   
$v_i, v'_i $ are the valuation and bid of buyer $i$, respectively, for a unit of the good, 
and $B_i$ is the budget of buyer $i$, i.e., the maximum total payment that $i$ can pay in the auction.
The valuation of each buyer is private information unknown to other buyers and the seller. 
Due to the impossibility theorem by Dobzinski et al. \cite{DLN2012}\footnote{They showed that there is no deterministic mechanism that satisfies all of incentive compatibility, individual rationality, and Pareto optimality if the budgets are private.}, we assume that the budget is public information available to the seller.
The seller determines the allocation based on a predetermined mechanism.

The allocation $\mathcal A:=(x, p)$ is a pair of $x\in \mathbb Z^N_+$ and $p\in \mathbb R^N$, where $x_i$ is the number of indivisible goods allocated to buyer $i$, and $p_i$ is the payment of buyer $i$ for their goods. 
Then, the budget constraints are described as $p_i\leq B_i\ (i\in N)$.
We are given an integer-valued monotone submodular function $f: 2^N\to \mathbb Z_+$ 
that represents the feasible allocation of goods.
Note that an integer-valued function 
$f: 2^N\to \mathbb Z_+$ is monotone submodular if it satisfies 
(i) $f(\emptyset)=0$, (ii) $f(S)\leq f(T)$ for each $S\subseteq T$, and (iii) 
$f(S\cup e)-f(S)\geq f(T\cup e)-f(T)$ for each $S\subseteq T$ and $e\in N\setminus T$.
For any set of buyers $S\subseteq N$, the buyers in $S$ can transact at most $f(S)$ amounts of goods through the auction. Note that $f(N)$ means the total goods sold in the auction.
This condition is equivalent to $x\in P(f)$ using the polymatroid
\[
P(f):=\{x\in \mathbb R^N_{+}\mid x(S)\leq f(S)\ (S\subseteq N)\}.
\]
We often denote $P(f)$ by $P$.
We also assume $f(N)=f(N\setminus i)$ for each $i\in N$, 
which implies that competition exists among buyers for each good at the beginning.

The utilities of the buyers are defined by 
	\[
	u_i(\mathcal A):=
	\begin{cases}
	\displaystyle 
	v_i x_i-p_i\quad\,  {\rm if}\ p_i\leq B_i, \\
	-\infty \qquad\quad  {\rm otherwise}
	\end{cases}
	\quad(i\in N).
	\]
Thus, the utilities of the buyers are quasi-linear if the budget constraints are satisfied, 
and otherwise, the utilities go to $-\infty$.
The utility of the seller is defined as the revenue of the seller, i.e., $u_s(\mathcal A):=\sum_{i\in N}p_i$.
The mechanism is a map $\mathcal M: \mathcal I\mapsto \mathcal A$ from information $\mathcal I$ to allocation $\mathcal A$. 
Note that $\mathcal I$ includes all information that the seller can access, 
and thus $\mathcal  I=(N, \{v'_i\}_{i\in N},\{B_i\}_{i\in N},f)$.
We call a mechanism $\mathcal M$ budget-feasible if, for any $\mathcal I$, mechanism $\mathcal M$ 
outputs an allocation that satisfies the budget constraints.

We consider to design an (economically) efficient budget-feasible 
mechanism $\mathcal M$ that satisfies incentive compatibility (IC) and individual rationality (IR).
A mechanism satisfies IC if for any $(\mathcal I, \{v_i\}_{i\in N})$, it holds 
$u_i(\mathcal M(\mathcal I_i))\geq  u_i(\mathcal M(\mathcal I))$ for each $i\in N$, where $\mathcal I_i$ denotes the information obtained from $\mathcal I$ by replacing $v'_i$ with $v_i$.
Intuitively, IC guarantees that it is the best strategy for each buyer is to report their true valuation.
A mechanism satisfies IR if, for each $i\in N$, there exists a bid $v'_i$ that guarantees her nonnegative utility.
When the mechanism satisfies IC, this means that for any $(\mathcal I, \{v_i\}_{i\in N})$, it holds $u_i(\mathcal M(\mathcal I_i))\geq 0$ for each $i\in N$.
Intuitively, IR guarantees that each buyer obtains nonnegative utility when the buyer reports her true valuation.

The efficiency of mechanisms is evaluated by the following measures:
A standard measure for auctions without budgets is social welfare (SW), which is 
defined as the sum of the valuations of the allocated goods for all buyers, and 
it can be interpreted as the sum of the utilities of all participants.
In other words, ${\rm SW}(\mathcal A):=\sum_{i\in N}v_i x_i=\sum_{i\in N}u_i (\mathcal A)+u_s(\mathcal A)$.
As mentioned, however, it is known (e.g., Dobzinski and Leme \cite{DL2014}, Lemma 2) that for any $\alpha<n$, there is no budget-feasible mechanism that achieves $\alpha$-approximation to the optimal SW with IC and IR.

An alternative measure for budget-constrained auctions is liquid welfare (LW), which is
defined by ${\rm LW}(\mathcal A):=\sum_{i\in N}\min(v_i x_i, B_i)$ for allocation $\mathcal A$.
LW represents the sum of the possible payments that buyers can pay for their allocated goods.

Another type of guarantee applicable to budget constraints is Pareto optimality (PO).
A mechanism satisfies PO if for any $(\mathcal I, \{v_i\}_{i\in N})$, there is no other allocation $\mathcal A'$ with (i) $u_i(\mathcal A')\geq u_i(\mathcal M(\mathcal I))$ for each $i\in N$, 
(ii) $u_s(\mathcal A')\geq u_s(\mathcal M(\mathcal I))$, and (iii) at least one inequality holds without equality.

\section{Polyhedral Clinching Auctions for Indivisible Goods}
In this section, we describe our mechanism. Our mechanism  
incorporates the polyhedral approach of Goel et al. \cite{GMP2015} 
to the (budgeted) clinching auctions in the previous indivisible settings  
 (e.g., \cite{BHLS2015, DLN2012, FLSS2011}).
The full description of our mechanism is presented in Algorithms 1 and 2.

	\begin{algorithm}[htb]
	\caption{Polyhedral\ Clinching\ Auction for Indivisible Goods} 
	\begin{algorithmic}[1]
	  \STATE $x_i=0,\ p_i:=0,\ d_i:=f(i)+1 \ 
	  (i\in N)\ {\rm and}\ c:=0$.\\
	  \WHILE{active buyers exist}
	  \STATE Increase $c$ until there appears an active buyer 
	  $j$ with $v'_{j}=c$ or $d_{j}=\frac{B_{j}-p_{j}}{c}$.
	  \WHILE{$\exists$ active buyer $j$ with $v'_{j}=c$}
	  \STATE Pick such a buyer $j$ and let $d_{j}:=0$.
	  \STATE Clinching$(f,x,p,d,c)$.
	  \ENDWHILE
	  \WHILE{$\exists$ active buyer $j$ with $d_j=\frac{B_j-p_j}{c}$}
	  \STATE Pick such a buyer $j$ and let $d_{j}:=d_{j}-1$.
	  \STATE Clinching$(f,x,p,d,c)$.
	  \ENDWHILE
	 \ENDWHILE
	 \STATE Output $(x^{\rm f}, p^{\rm f}):=(x,p)$.
	\end{algorithmic}
	\end{algorithm}

	\begin{algorithm}[htb]
	\caption{Clinching $(f,x,p,c,d)$} 
	\begin{algorithmic}[1]
	  \FOR{$i=1,2,\ldots,n$} 
	  \STATE 
   Let $\delta_i$ be the maximum non-negative integer with $P^i_{x,d}(\delta_{i})=P^i_{x,d}(0)$; see equations (\ref{feasible_transaction}) 
	  and~(\ref{clinching_definition}). Compute $\delta_i$.
    \STATE Clinch $\delta_i$; $\displaystyle x_i:=x_i+\delta_i,\ p_i:=p_i+c \delta_i,\ d_i:=d_i-\delta_i$.
	  \ENDFOR
	\end{algorithmic}
	\end{algorithm}
	
Now we outline the mechanism.
The price clock $c\in \mathbb R_{+}$ represents a transaction price for one unit of the good. 
Our mechanism is an ascending auction, where $c$ gradually increases.
For the current price $c$, the demand vector $d:=(d_i)_{i\in N}\in \mathbb Z^{N}_{+}$ is set so that $d_i$ represents the current demand of buyer $i$, while the initial value $d_i:=f(i)+1$ is an unavoidable technical requirement.\footnote{This is due to the following reason: If $d_i:=\infty$ (as in the divisible case), the number of iterations is not polynomially bounded. If $d_i:=f(i)$, buyer $i$ can drop out after clinching $f(i)$ units of the good with her budget remaining. Such a buyer may break the inequality of Theorem \ref{tightsets} (iii), which complicates the formulation of the tight sets lemma.}
A buyer $i$ is said to be {\it active} if $d_i>0$, and {\it dropping} if $d_i$ just reaches zero.	
In our mechanism, buyers can be dropping during the demand update in line 5 or 9 of Algorithm 1, or in line 3 of Algorithm 2.

Initially, the allocation $(x,p)$ is all zero, the price clock $c$ is set to zero, and the demand $d_i$ is set to $f(i)+1$ for each $i\in N$.
Our mechanism repeats the following procedure as long as active buyers exist: 
At the beginning of an iteration, the price clock $c$ is increased until 
there appears an active buyer $j$ with $c=v'_j$ or $d_j=(B_j-p_j)/c$. Then, the demands of all such buyers are updated.
After each of the update, 
the transactions of the buyers are computed in Algorithm 2.
If there is no active buyer, Algorithm 1 terminates and outputs the final allocation $(x^{\rm f},p^{\rm f})$.

The clinching steps in lines 6 and 10 are described in Algorithm 2.
We here consider two polytopes that represent the feasible transactions of buyers.
For the polymatroid $P$, and vectors $x\in P$ and $d\in \mathbb R^N_{+}$, 
we define the {\it remnant supply polytope} $P_{x,d}$ by
\begin{equation}
\label{remnant_supply_polytope}
P_{x,d}:=\{u\in \mathbb R^N_{+}\mid x+u\in P,\ u_i\leq d_i \ (i\in N)\}, 
\end{equation} 
which represents the feasible future transaction of buyers 
with demand $d$ given that $x$ has been clinched already.
In addition, for buyer $i$ and $w_i\leq d_i$, 
we define the polytope $P^i_{x,d}(w_i)$ by 
\begin{equation}
\label{feasible_transaction}
P^i_{x,d}(w_i):=\{u\vert_{N\setminus i}\mid  u\in P_{x,d}\ {\rm and}\ u_i=w_i\}, 
\end{equation}
which represents the feasible future transaction of other buyers $N\setminus i$ 
if buyer $i$ clinches $w_i$ unit of goods.
The polytopes $P_{x,d}$ and $P^{i}_{x,d}(w_i)$ are known to be polymatroids: 
The monotone submodular function $f_{x,d}$ for the polymatroid  
$P_{x,d}$ is given by 
\begin{align}
\label{naive}
f_{x,d}(S):&=\min_{S'\subseteq S}\{\min_{S''\supseteq S'}\{f(S'')-x(S'')\}+d(S\setminus S')\}\ \ (S\subseteq N)\nonumber\\
&=\min\Bigl(f_{x,d}(S\setminus k)+d_k, \min_{S'\subseteq S\setminus k}\{\min_{S''\supseteq S'}
\{f(S''\cup k)-x(S''\cup k)\}+d(S\setminus (S'\cup k)\}\Bigr)\\
&\qquad\qquad\qquad\qquad\qquad\qquad\qquad\qquad
\qquad\qquad\qquad\qquad\qquad (S\subseteq N, k\in S)\nonumber.
\end{align} 
See Section 3.1 of Fujishige \cite{F2005}. 
The second equation is noted here because it is often used in technical discussions in Section 4 and beyond.

The clinching amount $\delta_i$ of buyer $i$ is then defined by 
\begin{equation}
\label{clinching_definition}
\delta_i:=\sup\{w_i\geq 0\mid P^i_{x',d'}(w_i)=P^i_{x',d'}(0)\},
\end{equation}
where $x'$ and $d'$ denote the allocation of goods and the demand vector, respectively, just before the clinching step of buyer $i$ in Algorithm 2. 
This means that each buyer $i$ clinches the maximal possible amount $\delta_i$, not affecting the feasible transactions of other buyers $N\setminus i$. 
This intuition of clinching is consistent with the ones in previous works
\cite{A2004, BCMX2010,BSV2011, BHLS2015,DHS2015, DLN2012, FLSS2011,GMP2014,GMP2015, GMP2020} on clinching auctions.
Moreover, let $x$ and $d$ be the allocation of goods and the demand vector, respectively, just before Algorithm~2.
Then, $\delta:=(\delta_i)_{i\in N}$ can also be computed via $f_{x,d}$:

\begin{lemma}
\label{clinch_amount}
In Algorithm 2, it holds $\delta_i=f_{x,d}(N)-f_{x,d}(N\setminus i)$ for each~$i\in N$.
\end{lemma}
Lemma \ref{clinch_amount} implies that the amount of 
goods allocated to each buyer in Algorithm 2 is independent of the order of the buyers. The proof is given in Appendix B, as the proof for divisible setting
(Goel et al. \cite[Lemma 3.5]{GMP2015})
can be easily adapted.

In the following, we investigate the properties of our mechanism. 
We first consider the properties specific to our indivisible setting: 
There are two major differences between our mechanism and the one in Goel et al.  \cite{GMP2015}.
The first is integrality of buyer's demand due to the indivisibility: 
By (\ref{naive}) and the assumption that $f$ is integer-valued, 
if $x$ and $d$ are integer vectors, 
the function $f_{x,d}$ is also integer-valued.
In that case, by Lemma \ref{clinch_amount}, 
$\delta$ is an integer vector in Algorithm 2 and 
thus $x$ and $d$ remain to be integer vectors.
Since $x$ and $d$ are integer vectors at the beginning of the auction, 
we have the following: 

\begin{lemma}
\label{integer-clinching}
Throughout the auction, $x,d$, $\delta$ are all integer vectors.
\end{lemma}

The second is on price update: 
Our mechanism sets a common price $c$ for all buyers and does not use a fixed step size for price increases.
This is based on the idea of Bikhchandani et al. \cite{BSV2011} and Fiat et al. \cite{FLSS2011}, 
and it plays an essential role 
in the iteration bounds, particularly, in the computational complexity.
In our mechanism, the total sum of initial demands is $\sum_{i\in N} f(i)+n$, 
and it is guaranteed that the total sum of the demands is decreased by at least one
per iteration.
Our mechanism terminates after at most $\sum_{i\in N} f(i)+n$ iterations, 
which is bounded by $n(f(N)+1)$ based on the monotonicity of $f$.

Moreover, each iteration can be computed in polynomial time.
In line 3 of Algorithm 1, the price is updated to $\min_{i\in N: d_i>0}\min\{v'_i, (B_i-p_i)/d_i\}$, which can be computed in polynomial time.
By Lemma \ref{clinch_amount}, 
Algorithm 2 can also be executed in polynomial time 
by a submodular minimization algorithm 
(e.g., Lee et al. \cite{LSW2015}), 
provided the value oracle of $f$ is given.
Therefore, we have the following:\footnote{If the number of goods $f(N)$ is given in the binary representation, it is pseudo-polynomial. Such a model is applied for the case where $f(N)$ is assumed to be large; e.g., Dobzinski and Nisan \cite{DN2010}. On the other hand, it is also natural to regard a large amount of goods as being divisible. Therefore, we do not make such an assumption, and consider the number $f(N)$ as a part of the input size.} 

\begin{theorem}
Our mechanism can be implemented in polynomial time.
\end{theorem}

	Next, we establish the budget feasibility of our mechanism: 
	\begin{theorem}
	\label{BF}
	Our mechanism is budget-feasible.
	\end{theorem}
	To prove Theorem \ref{BF}, we use the following lemma also used in Sections 4 and 5.
	\begin{lemma}
	\label{d}
Consider a step of an iteration in Algorithm~1. For an active buyer $i$, it holds 
\begin{itemize}
\item[(i)] If the demand $d_i$ has never been updated in line 9 so far, then $d_i=f(i)+1-x_i\leq  (B_i-p_i)/c$. 
\item[(ii)] If the demand $d_i$ has just been updated in line 9, 
then $d_i= (B_i-p_i)/c-1$ for the rest in this iteration. 
\item[(iii)] In other cases, it holds $d_i=\left\lfloor  (B_i-p_i)/c\right\rfloor$.
\end{itemize}
	\end{lemma}
	\begin{proof}
    At the beginning of the auction, by line 1 of Algorithm 1, 
    it holds $d_i=f(i)+1-x_i\leq  (B_i-p_i)/c$.
    In the following ($c>0$), we use the property that 
    all of $d_i$, $(B_{i}-p_{i})/c$, $\left\lfloor(B_{i}-p_{i})/c\right\rfloor$ decrease by the same amount $\delta_i$ in Algorithm 2.
    This property immediately holds by the integrality of $\delta_i$ (Lemma \ref{integer-clinching}).
    
    Suppose that the demand $d_i$ has never been updated in line 9 so far. 
	Then, the demand $d_i$ was decreased only by clinching in Algorithm~2. 
	By the above property, $d_i$ and $(B_{i}-p_{i})/c$ were decreased 
    by $\delta_i$.
	Therefore, we have $d_i=f(i)+1-x_i\leq  (B_i-p_i)/c$. 
	
	If the demand $d_i$ has just been updated in line 9, 
	it holds $d_i=(B_{i}-p_{i})/c-1$ after the update.
    By the above property, in the rest in this iteration, 
	the demand $d_i$ is changed with keeping 
    $d_i=(B_{i}-p_{i})/c-1$.

    The remaining case consists of buyers whose demand 
    has been updated in line 9 of previous iterations  
    and has not yet been updated in this iteration.
    Then, after the first demand update, 
	the price update in line 3 is performed with keeping the equality $d_i=\left\lfloor(B_{i}-p_{i})/c\right\rfloor$. 
    By the above property, 
    this equation is preserved after the execution of Algorithm 2.
	\end{proof}
	\begin{proof}[Proof of Theorem \ref{BF}]
	We consider the case where an active buyer $i$ is dropping in an iteration.
	Let $c$ be the price clock in the iteration, and  
	$p$ and $d$ be the payment vector and the demand vector, respectively, 
	just before the dropping of buyer $i$.
	Then, Lemma \ref{d} implies that $B_i-p_i\geq c d_i>0$. 
	If buyer $i$ is dropping during the demand update in line 5 or 9, 
	then we have $B_i-p^{\rm f}_i=B_i-p_i\geq c d_i>0$.
	Suppose that buyer $i$ is dropping in Algorithm 2. 
	Then, in any case of Lemma \ref{clinch_amount}, 
    it holds $\delta_i=f_{x,d}(N)-f_{x,d}(N\setminus i)$. Then, by (\ref{naive}), it holds 
    $f_{x,d}(N)\leq f_{x,d}(N\setminus i)+d_i$,
    which means $\delta_i\leq d_i$.
    Then, we have 
    $B_i-p^{\rm f}_i=B_i-p_i-c\delta_i\geq B_i-p_i-c d_i\geq 0$.
	Therefore, our mechanism is budget-feasible. 
	\end{proof}

Moreover, our mechanism inherits several desirable properties from 
the polyhedral clinching auction by Goel et al. \cite{GMP2015}: 
\begin{theorem}
	\label{IC_IR}
	Our mechanism satisfies IC and IR.
\end{theorem}	
\begin{proof}
	For each buyer $i$, the bid $v'_i$ is used to only determine when the buyer $i$ 
	is dropping. If $v'_i<v_i$, buyer $i$ may miss some goods at a price below their valuation.
	In addition, if $v'_i>v_i$, buyer $i$ may clinch some goods at a price greater than their valuation. Then, truthful bidding is 
	the best strategy for buyers, and therefore our mechanism satisfies IC. 
	The mechanism satisfies IR because each buyer clinches some goods at a price lower than the buyer's valuation if $v'_i=v_i$. 
\end{proof}

From Theorem \ref{IC_IR}, our mechanism satisfies IC, and therefore, 
we assume in the rest of the paper that all buyers bid truthfully, that is, $v'_i=v_i$ for every $i\in N$. 

\begin{theorem}
\label{all_goods}
All goods are sold in our mechanism, i.e., $x^{\rm f}(N)=f(N)$.
\end{theorem}

In the proof, we utilize the properties of the monotone submodular function $f_{x,d}$ in Algorithm~1, which will also be used in the subsequent~sections.

\begin{lemma}
\label{fxd_properties}
The following holds:
\begin{itemize}
\item[(i)] At the beginning of the auction, it holds $f_{x,d}(S)=f(S)$ for each $S\subseteq N$.
\item[(ii)] Throughout the auction, it holds $\displaystyle f_{x,d}(S)=\min_{S'\subseteq S}\{f(S')-x(S')+d(S\setminus S')\}$
 for each $S\subseteq N$.
\item[(iii)] By the execution of Algorithm~2, the value $x(S)+f_{x,d}(S)$ is unchanged for each $S\subseteq N$.
\item[(iv)] Just before the demand update in line 5 or 9, it holds $f_{x,d}(N)=f_{x,d}(N\setminus i)$ for each $i\in N$.
\item[(v)] During the demand update in line 5 or 9, the value $f_{x,d}(N)$ is unchanged.
\end{itemize}
\end{lemma}

\begin{proof}
	(i) At the beginning of the auction, by $x_i=0$ 
    and $d_i=f(i)+1\ (i\in N)$, we see from (\ref{remnant_supply_polytope}) that $P=P_{x,d}$.
    Since two polymatroids are equal if and only if the corresponding monotone submodular functions are equal, 
    we have $f_{x,d}(S)=f(S)$.

(ii) This property is the same as Proposition 3.8 of Sato \cite{S2023} for his divisible setting. His proof can be easily adapted in our setting.
The proof is given in Appendix B.

(iii) From (ii), we have 
\begin{equation}
\label{x+f_xd}
x(S)+f_{x,d}(S)=\min_{S'\subseteq S}\{f(S')+x(S\setminus S')+d(S\setminus S')\}
\ \ (S\subseteq N)
\end{equation}
just before Algorithm~2. 
Suppose that buyers in $S$ clinch $\delta(S)$ amounts of goods 
in Algorithm~2.
For each $i\in S$, it holds that $x_i$ increases by $\delta_i$ and $d_i$ decreases by $\delta_i$, and thus $x_i+d_i$ is unchanged.
Since (\ref{x+f_xd}) holds for $x$ and $d$ after the execution of Algorithm~2, the value $x(S)+f_{x,d}(S)$ is unchanged. 

	(iv) 
        At the beginning of the auction, by (i) and $x_i=0\ (i\in N)$, we have $x(N)+f_{x,d}(N)=f(N)$ and 
    $x(N\setminus i)+f_{x,d}(N\setminus i)=f(N\setminus i)$ for each $i\in N$.
	Then, by the assumption $f(N)=f(N\setminus i)$, we have 
	$f_{x,d}(N\setminus i)=f_{x,d}(N)$.

        Now consider the moment just after Algorithm 2 has finished.
	By Lemma \ref{clinch_amount}, in Algorithm~2,
	it holds $\delta_i=f_{x,d}(N)-f_{x,d}(N\setminus i)$.
	Then, we have 
	$f_{x,d}(N)-\delta(N)=f_{x,d}(N)-\delta_i-
	\delta(N\setminus i)=f_{x,d}(N\setminus i)-\delta(N\setminus i)$.
	By (iii), $f_{x,d}(N)-\delta(N)$ and $f_{x,d}(N\setminus i)-\delta(N\setminus i)$ 
	are equal to $f_{x,d}(N)$ and $f_{x,d}(N\setminus i)$, respectively, 
	just after Algorithm 2 has finished. Then, we also have
        $f_{x,d}(N)=f_{x,d}(N\setminus i)$.
 
	Therefore, it holds $f_{x,d}(N)=f_{x,d}(N\setminus i)$ 
        just before the demand update in line 5 or~9.
 
	(v) Just before the demand update of $i$, by (iv), 
        it holds $f_{x,d}(N)=f_{x,d}(N\setminus i)$.
	After the demand update, $f_{x,d}(N\setminus i)$ is unchanged 
    since it is independent of $d_i$, 
    while $f_{x,d}(N)$ is non-increasing by (ii).
    Thus, we have $f_{x,d}(N)\leq f_{x,d}(N\setminus i)$.
    By the monotonicity of $f_{x,d}$, 
    we also have $f_{x,d}(N)\geq f_{x,d}(N\setminus i)$.
	Therefore, $f_{x,d}(N) (=f_{x,d}(N\setminus i))$ is also unchanged.
\end{proof}

\begin{proof}[Proof of Theorem \ref{all_goods}]
We show that $x(N)+f_{x,d}(N)=f(N)$ holds throughout Algorithm 1.
Then, at the end of the auction, by Lemma \ref{fxd_properties} (ii) 
and $d_i=0\ (i\in N)$, it holds $(0\leq) f_{x,d}(N)\leq d(N)=0$,
from which we have $x^{\rm f}(N)=f(N)$.

From Lemma \ref{fxd_properties} (i) and $x_i=0\ (i\in N)$, 
at the beginning of the auction, 
it holds $x(N)+f_{x,d}(N)=f(N)$.
From Lemma \ref{fxd_properties} (v), during the demand update in line 5 or 9,
$x(N)$ and $f_{x,d}(N)$ are unchanged.
Moreover, from Lemma \ref{fxd_properties} (iii), 
by the execution of Algorithm 2, 
$x(N)+f_{x,d}(N)$ is unchanged.
Therefore, we have $x(N)+f_{x,d}(N)=f(N)$ throughout Algorithm 1.
\end{proof}

\section{Tight Sets Lemma}	 
In this section, we establish the {\it tight sets lemma} for our mechanism.
It characterizes the structure of the final allocation 
according to the dropping situation of buyers, 
and will be the basis for efficiency guarantees in Section 5.

Consider an execution of Algorithm 1 for the input $\mathcal I$. Remark that, as mentioned just after Theorem \ref{IC_IR}, we assume that all buyers report their bids truthfully, i.e., $v'_i=v_i$ for every $i\in N$. To analyze the dropping situation of buyers, as in Goel et al. \cite{GMP2014} for divisible case, we make the following definition:
	\begin{definition}[Goel et al. \cite{GMP2014}]
        \label{dropping_prices}
        Given an execution of Algorithm 1:
        \begin{itemize}
        \item The {\it dropping price} $c^{\rm f}_i$ of buyer $i$ is the price in which $i$ is dropping. Obviously, it holds $v_i\geq c^{\rm f}_i\  (i\in N)$. 
        \item Let $i_1, i_2,\ldots, i_t$ be the buyers dropping during demand update in line 5 or 9, where they are sorted in the {\it reverse} order of their dropping, i.e., $c^{\rm f}_{i_1}\geq \cdots\geq c^{\rm f}_{i_t}$. 
        If $i_k$ is dropping in line 5, then it holds $c^{\rm f}_{i_k}=v_{i_k}$. If $i_k$ is dropping in line 9, then it holds $c^{\rm f}_{i_k}<v_{i_k}$. 
        \item For each $k\in [t]$, let $X_k$ denote the set of active buyers just before the dropping of $i_k$. Obviously, it holds $i_k\in X_k \setminus X_{k-1}$ and $\emptyset = X_0 \subset X_1 \subset X_2 \subset \cdots \subset X_t = N$. 
         \item A set of buyers $S\subseteq N$ is {\it tight} if it holds $x^{\rm f}(S)=f(S)$.  
        \end{itemize}
        \end{definition}
    The dropping of buyers 
    in Algorithm 1 is illustrated in Figure \ref{image_tightsets}.
    Now we state the main result in this section. 
    The final allocation of our mechanism can be summarized as follows:

    \begin{figure}[htbp]
	\begin{center}
	\includegraphics[width=113mm]{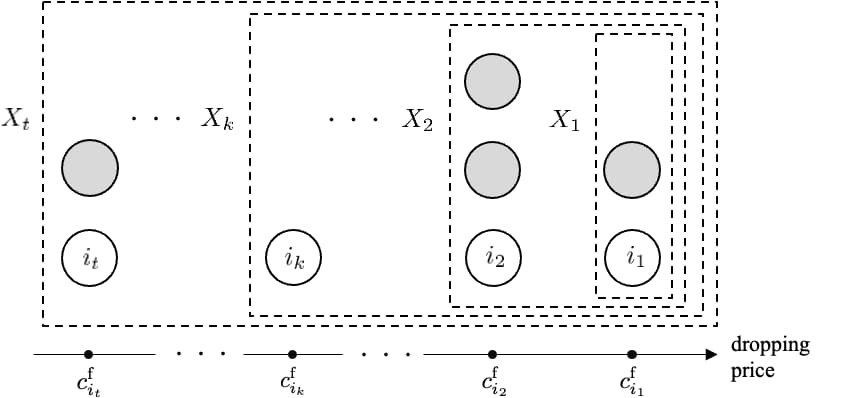}
    \caption{Illustration of the dropping of buyers in Algorithm 1. The white circles represent buyers $i_1, i_2,\ldots, i_t$ and the gray shaded circles represent other buyers. For each $k\in [t]$, the dropping price $c^{\rm f}_i$ of buyer $i\in X_k\setminus X_{k-1}$ is equal to $c^{\rm f}_{i_k}$, as shown in Theorem \ref{tightsets} (ii).}
	\label{image_tightsets}
	\end{center}
    \end{figure}

	\begin{theorem}[Tight sets lemma]
        \label{tightsets}
        For each $k\in [t]$, the following holds:
	\begin{itemize}
        \item[{\rm (i)}]  $X_k$ is tight, i.e., it holds $x^{\rm f}(X_k)=f(X_k)$.
        \item[{\rm (ii)}] For each $i \in X_k\setminus X_{k-1}$, it holds $c^{\rm f}_i =c^{\rm f}_{i_k}$.
	\item[{\rm (iii)}] For buyer $i \in X_k\setminus (X_{k-1} \cup i_k)$, it holds $(B_i-p^{\rm f}_i)/c^{\rm f}_i\leq 1$.
    \item[{\rm (iv)}] 
    For the following properties, it holds $(a)\Rightarrow (b)\Rightarrow (c)$.
    \begin{itemize}
    \item[(a)] There exists a buyer $\ell \in X_k\setminus (X_{k-1}\cup i_k)$ with $(B_{\ell}-p^{\rm f}_{\ell})/c^{\rm f}_{\ell}=1$.
    \item[(b)] Buyer $i_k$ is dropping in line 9, i.e., $c^{\rm f}_{i_k}<v_{i_k}$.
    \item[(c)] It holds $(B_{i_k}-p^{\rm f}_{i_k})/c^{\rm f}_{i_k}=1$ and $c^{\rm f}_{i_k}<v_i$ for each $i \in X_k$.
    \end{itemize}
        \end{itemize}
	\end{theorem}

Due to the indivisibility, our tight sets lemma is more complicated than that in Goel et al. \cite{GMP2015}, in which the above properties (iii) and (iv) are simplified into one condition: It holds $c^{\rm f}_{i_k}=v_{i_k}$ and $p^{\rm f}_i=B_i$ for each $i \in X_k\setminus (X_{k-1} \cup i_k)$. In addition, the property (ii) is needed for our efficiency guarantees, as in the concave budget setting in Goel et al. \cite{GMP2014}.

    Theorem \ref{tightsets} is a consequence of the following proposition.
    
\begin{proposition}
\label{dropping_of_buyers}
    The following properties hold:
    \begin{itemize}
    \item[(i)]
    	If a buyer $i$ is dropping in line 3 of Algorithm 2, then there is $k\in [t]$ such that buyer $i_k$ was dropping in line 5 or 9 just before this execution of Algorithm~2.
    \item[(ii)]
	Just before the demand update in line 5 or 9, it holds $x^{\rm f}(S) = f(S)$, where $S$ is the set of active buyers.

    \end{itemize}
    \end{proposition}
    
Assuming this, we first complete the proof of the tight sets lemma. 

\begin{proof}[Proof of Theorem \ref{tightsets}]
(i) The tightness of $X_k$ immediately holds by Proposition \ref{dropping_of_buyers} (ii).

(ii) By the definition of $X_k$, 
     each buyer $i\in X_k\setminus (X_{k-1}\cup i_k)$ is dropping in Algorithm~2.
     Then, by Proposition \ref{dropping_of_buyers} (i), 
     the buyer $i$ has the same dropping price as $i_k$.

(iii) 
Let $i\in X_{k}\setminus (X_{k-1}\cup i_k)$.
Suppose that buyer $i$'s demand has never been updated in lines 5 and~9 before her dropping.
Then, by Lemma \ref{d} (i) and the polymatroid constraint $x^{\rm f}_i\leq f(i)$, 
we have $d_i=f(i)+1-x^{\rm f}_i>0$, which means that $i$ must be always active. Thus, such a way of dropping does not happen.
Therefore, the demand $d_i$ has been updated in line 9 at least once before her dropping. 
By Lemma~\ref{d} (ii) and (iii), after the first demand update, $d_i$ changes with keeping the inequality $d_i\geq  (B_i-p_i)/c-1$.
Just when the demand $d_i$ decreased to zero by Algorithm 2, 
then $d_i$ and $(B_i-p_i)/c^{\rm f}_i$ decreased by the same amount $\delta_i$.
This implies $0\geq (B_i-p^{\rm f}_i)/c^{\rm f}_i-1$.

(iv) 
(a)$\Rightarrow$(b): If there exists a buyer $\ell \in X_k\setminus (X_{k-1}\cup i_k)$ with $(B_{\ell}-p^{\rm f}_{\ell})/c^{\rm f}_{\ell}=1$, then it holds $d_{\ell}=(B_{\ell}-p_{\ell})/c^{\rm f}_{\ell}-1\in \mathbb Z_+$ just before Algorithm 2 where $\ell$ dropped (since $d_\ell$ and $(B_\ell-p_\ell)/c_\ell$ decreases by the same quantity).
By the proof of (iii), the demand $d_\ell$ had been updated in line 9 at least once before her dropping. Then, by Lemma \ref{d} (ii) and (iii) and $\left\lfloor (B_{\ell}-p_{\ell})/c^{\rm f}_{\ell}\right\rfloor>(B_{\ell}-p_{\ell})/c^{\rm f}_{\ell}-1$, it necessarily holds that $\ell$'s demand had been updated in line 9 in the same iteration before her dropping.
Therefore, $\ell$ was dropped in line 10 of Algorithm~1 and by Proposition \ref{dropping_of_buyers} (i), 
$i_{k}$ had been dropped in line 9.

(b)$\Rightarrow$(c):
Suppose that $i_{k}$ was dropped in line 9. 
Since the demand $d_{i_k}$ of buyer $i_k$ decreased by one just before her dropping, it holds $(B_{i_{k}}-p_{i_{k}})/c^{\rm f}_{i_{k}}=1$.
Moreover, by the price update and the demand update in line 5, all buyers with valuations lower than or equal to $c^{\rm f}_{i_k}$ had already been dropped before the demand update of $i_k$.
Therefore, it holds $c^{\rm f}_{i_k}<v_i$ for each $i\in X_k$.
\end{proof}

The remainder of this section is devoted to proving Proposition \ref{dropping_of_buyers}.
We first introduce {\it unsaturation}, a binary relation among buyers in Section 4.1, and then prove Proposition \ref{dropping_of_buyers} in Section~4.2.

\subsection{Unsaturation Relation}

Here we focus on a fixed iteration.
Also, for the demand vector $d\in \mathbb Z_+^N$, we define $d^{-k}$ by $(d^{-k})_k := 0$ and $(d^{-k})_i := d_i$ for $i \neq k$.
Now we use the following binary relation: \footnote{To simplify the arguments, we use the equivalent property in \cite[Lemma 4.13]{GMP2014} instead of the original definition.}

\begin{definition}[Goel et al. \cite{GMP2014}]
\label{unsaturation}
For buyers $i,k$, we denote $i\preceq k$ if 
$f_{x,d^{-k}}(N)-f_{x,d^{-k}}(N\setminus i)=(d^{-k})_i$.
\end{definition}

This binary relation is closely related to the demands of buyers and the clinching amount.
In the proof of Proposition \ref{dropping_of_buyers}, we use the following properties:
\begin{lemma}
\label{unsaturation_property}
The following properties hold:
\begin{itemize}
    \item[(i)] $i\preceq i$, i.e., $\preceq$ is reflexive.
    \item[(ii)] If $i\preceq k$, then $f_{x,d^{-k}}(S)-f_{x,d^{-k}}(S\setminus i)=(d^{-k})_i$ for any $S\ni i$.
    \item[(iii)]  Just before line 5 or 9, if $i\preceq k$, then it holds $d_i\leq d_k$.
    \item[(iv)]Suppose that the demand $d_i$ of buyer $i$ decreases by $\Delta_i$ in line 5 or 9.
	For each buyer $k\neq i$, 
	if $i\preceq k$ just before the demand update, then it holds $\delta_k =\Delta_i$ in the subsequent  Algorithm~2.
	Otherwise, $\delta_k<\Delta_i$.
\end{itemize}
\end{lemma}

To prove this, we use
\begin{equation}
\label{fact}
f_{x,d}(S\setminus k)=f_{x,d^{-k}}(S\setminus k)\quad (S\ni k), 
\end{equation}
which obviously holds because $f_{x,d}(S\setminus k)$ 
and $f_{x,d^{-k}}(S\setminus k)$ is independent of $d_k$:

\begin{proof}
    (i) (ii) These properties are shown in \cite[Lemmas 4.17 and 4.15]{GMP2014}, respectively, for their divisible setting. Their proof can be easily adapted in our setting. 
    
    (iii) By (i) and (\ref{fact}), it holds 
        $f_{x,d^{-j}}(N)=f_{x,d^{-j}}(N\setminus j)=f_{x,d}(N\setminus j)$.
        Combining this with Lemma~\ref{fxd_properties}~(iv),
        just before line 5 or 9, 
        it holds $f_{x,d}(N)=f_{x,d}(N\setminus j)=f_{x,d^{-j}}(N)$ 
        for each $j\in N$.
        Then, substituting $i$ and $k$ for $j$, we have ($\ast$1) 
        $f_{x,d}(N\setminus i)=f_{x,d}(N)=f_{x,d^{-k}}(N)$ just before line~5~or~9.

	By (\ref{naive}) and (\ref{fact}), it holds ($\ast$2)
	$f_{x,d}(N\setminus i)\leq f_{x,d}(N\setminus \{i,k\})+d_k=f_{x,d^{-k}}(N\setminus \{i,k\})+d_k$.
        Applying (ii) for $i\preceq k$ and $k\preceq k$, 
	we have ($\ast$3)$f_{x,d^{-k}}(N)=f_{x,d^{-k}}(N\setminus \{i,k\})+(d^{-k})_i+(d^{-k})_k=f_{x,d^{-k}}(N\setminus \{i,k\})+d_i$.
	By ($\ast$1), ($\ast$2), and ($\ast$3), we have 
    \[
f_{x,d^{-k}}(N\setminus \{i,k\})+d_k\geq f_{x,d}(N\setminus i)=f_{x,d^{-k}}(N)=f_{x,d^{-k}}(N\setminus \{i,k\})+d_i.    
    \]
	Therefore, if $i\preceq k$ just before line 5 or 9, then it holds $d_i\leq d_k$.

    (iv) Let $d'$ be the demand vector after the demand update of $i$, 
        i.e., the vector obtained from $d$ by replacing $d_i$ with $d_i-\Delta_i$.
        Then, by Lemma \ref{fxd_properties} (iv) and (v), it holds 
	$f_{x,d'}(N)=f_{x,d}(N)=f_{x,d}(N\setminus k)$.
        We utilize Lemma \ref{clinch_amount}: 
        $\delta_k=f_{x,d'}(N)-f_{x,d'}(N\setminus k)$.
        We show that $f_{x,d'}(N)-f_{x,d'}(N\setminus k)=\Delta_i$ if $i\preceq k$, and $f_{x,d'}(N)-f_{x,d'}(N\setminus k)<\Delta_i$ otherwise. To this end, we examine how $f_{x,d'}(N\setminus k)$ can be represented.

        First, we show  
        \begin{equation}
        f_{x,d'}(N\setminus k)=\min\bigl(f_{x,d}(N\setminus k),f_{x,d}(N\setminus \{i,k\})+d'_i\bigr).
        \end{equation}
        Define a constant $F$ by 
        \[
        F:=\min_{S'\subseteq N\setminus \{i,k\}}\{\min_{S''\supseteq S'}
\{f(S''\cup i)-x(S''\cup i)\}+d(N\setminus (S'\cup \{i,k\})\}.
        \]
        Then, by (\ref{naive}), we have 
	\begin{align*}
	&f_{x,d}(N\setminus k)
        =\min\bigl(f_{x,d}(N\setminus \{i,k\})+d_i,F\bigr),\\
        &f_{x,d'}(N\setminus k)=\min\bigl(f_{x,d'}(N\setminus \{i,k\})+d'_i,F\bigr)
        =\min\bigl(f_{x,d}(N\setminus \{i,k\})+d'_i,F\bigr),
	\end{align*}
        where the last equality holds since $f_{x,d}(N\setminus \{i,k\})$ is independent of $d_i$.

        If $F\leq f_{x,d}(N\setminus \{i,k\})+d'_i$, by $d'_i<d_i$, 
        we have $f_{x,d'}(N\setminus k)=f_{x,d}(N\setminus k)=F\leq f_{x,d}(N\setminus \{i,k\})+d'_i<f_{x,d}(N\setminus \{i,k\})+d_i$.
        Otherwise, we have 
        $f_{x,d'}(N\setminus k)=f_{x,d}(N\setminus \{i,k\})+d'_i<\min\bigl( f_{x,d}(N\setminus \{i,k\})+d_i,F\bigr)=f_{x,d}(N\setminus k)$.
        Therefore, we have 
        \begin{equation}
        \label{Nsetminusk}
        f_{x,d'}(N\setminus k)=\min\bigl(f_{x,d}(N\setminus k),f_{x,d}(N\setminus \{i,k\})+d'_i\bigr).
        \end{equation}

        Next we focus on the relationship between $f_{x,d}(N\setminus k)$ and $f_{x,d}(N\setminus \{i,k\})$.
        Suppose that $i\preceq k$. Then, we have 
	\begin{align}
        \label{unsaturation_eq}
	f_{x,d}(N\setminus \{i,k\})&=f_{x,d^{-k}}(N\setminus \{i,k\})
        =f_{x,d^{-k}}(N\setminus i) \nonumber\\ 
        &=f_{x,d^{-k}}(N)-d_i
        =f_{x,d^{-k}}(N\setminus k)-d_i
	=f_{x,d}(N\setminus k)-d_i, 
	\end{align}
        where the first and last equalities hold by (\ref{fact}), 
        and the second and fourth equalities hold by (i) ($k\preceq k$), 
        and the third equality holds by $i\preceq k$.
        By (\ref{Nsetminusk}), this means that $f_{x,d'}(N\setminus k)=\min\bigl(f_{x,d}(N\setminus k),f_{x,d}(N\setminus \{i,k\})+d'_i\bigr)=f_{x,d}(N\setminus k)+d'_i-d_i=f_{x,d}(N\setminus k)-\Delta_i$.
        By Lemma \ref{clinch_amount}, we have 
        $\delta_k=f_{x,d'}(N)-f_{x,d'}(N\setminus k)=f_{x,d}(N\setminus k)-\bigl(f_{x,d}(N\setminus k)-\Delta_i\bigr)=\Delta_i$.

	Suppose that $i\npreceq k$. By Definition \ref{unsaturation} and (\ref{naive}) for $f_{x,d^{-k}}(N)$, 
        the third equation 
        in (\ref{unsaturation_eq}) changes to ``$>$''.
        Then, we have 
	$f_{x,d}(N\setminus \{i,k\})>f_{x,d}(N\setminus k)-d_i$, 
        which means $f_{x,d'}(N\setminus k)>f_{x,d}(N\setminus k)-\Delta_i$.
        By Lemma \ref{clinch_amount}, we have 
        $\delta_k=f_{x,d'}(N)-f_{x,d'}(N\setminus k)<f_{x,d}(N\setminus k)-\bigl(f_{x,d}(N\setminus k)-\Delta_i\bigr)=\Delta_i$.
\end{proof}

\subsection{Proof of Proposition \ref{dropping_of_buyers}}
Using Lemma \ref{unsaturation_property}, we prove Proposition \ref{dropping_of_buyers}. 
Note that we use the similar arguments as in Goel et al. \cite[Lemmas 4.3 and 4.4]{GMP2014}.

	\begin{proof}[Proof of Proposition \ref{dropping_of_buyers}] 
	(i) We show the contraposition: After the demand update of a buyer, 
        if the buyer still has a positive demand, then no buyer is dropping in the subsequent execution of Algorithm~2.
        Since the buyer is dropping after the update in line 5, 
        it suffices to consider the demand update in line~9.
 
        Let $d$ (resp. $d'$) denote the demand vector, 
	just before (resp. after) the demand update of $i$
        in line~9, i.e., $d'_i=d_i-1>0$ and $d'_j=d_j\ (j\in N\setminus i)$. 
        Then, we have 
        \[
        f_{x,d'}(N)=f_{x,d}(N)=f_{x,d}(N\setminus i)=f_{x,d'}(N\setminus i),
        \]
        where the first (resp. the second) equality holds by  
	Lemma \ref{fxd_properties} (v) (resp. (iv)) and the third 
        equality holds since $f_{x,d}(N\setminus i)$ is independent of $d_i$.
	By Lemma \ref{clinch_amount}, 
	we have $\delta_i=f_{x,d'}(N)-f_{x,d'}(N\setminus i)=0$.
    Then, $i$ is still active just after Algorithm 2 has finished.
    
	Now we consider the clinching amount $\delta_k$ of buyer $k\neq i$ in Algorithm 2.
	If $i\preceq k$ just before line 9, it follows from Lemma \ref{unsaturation_property} (iii) that $d_i\leq d_k$.
	Thus, by Lemma \ref{unsaturation_property} (iv) and $d'_i>0$,
    we have $\delta_k=d_i-d'_i\leq d_k-d'_i<d_k$.
	Otherwise, by Lemma \ref{unsaturation_property} (iv),
    we have $\delta_k<d_i-d'_i=1$.
     By Lemma~\ref{integer-clinching}, this means $\delta_k=0$.
    In both cases, buyer $k$ is still active just after Algorithm 2 has finished.

	(ii) We show by induction. At the beginning of the auction, since $x^{\rm f}(N)=f(N)$ by Theorem~\ref{all_goods}, 
    the set of active buyers is tight.
	Consider a moment just before line 5 or 9 
        in an iteration, 
        where $T$ is the set of active buyers.
        Suppose that $x^{\rm f} (T)=f(T)$.
	As long as the number of active buyers remains unchanged, 
        the claim holds trivially. 
	Then, by (i), 
        it suffices to consider the case where there exists a buyer $k$ dropping in line 5 or 9.
	Let $x$ and $d$ denote the allocation and the demand vector, 
	respectively, just before the demand update of $k$.
        Then, $d^{-k}$ represents the demand vector just after the update.
	For buyer $i$ ($i\neq k$), if $i\preceq k$ just before the update, 
	by Lemma~\ref{clinch_amount}, we have 
	$\delta_i=f_{x,d^{-k}}(N)-f_{x,d^{-k}}(N\setminus i)
        =(d^{-k})_i=d_i$.
        This means that these buyers are dropping by Algorithm~2.
	Let $S$ denote the set of buyers $i\in T$ with $i\npreceq k$.
	Obviously, it holds $k\notin S$ by $k\preceq k$ 
        (Lemma~\ref{unsaturation_property}~(i)).
	By Lemma~\ref{clinch_amount}, 
    it holds $\delta_j=f_{x,d^{-k}}(N)-f_{x,d^{-k}}(N\setminus j)
        <(d^{-k})_j=d_j$ for each $j\in S$. Thus, 
	only the buyers in $S$ are still active just after Algorithm 2 has finished.
	Then, it remains to show that $x^{\rm f} (S)=f(S)$, 
	which is obtained by 
	$f_{x,d}(S)=f(S)-x(S)$ and $f_{x,d}(S)=x^{\rm f}(S)-x(S)$.

    In the following, we fix the above sets $S$ and $T$.
	Firstly, we show that $f_{x,d}(S)=x^{\rm f}(S)-x(S)$.
    Let $x'$ and $d'$ denote the allocation and the demand vector, 
	respectively, just after Algorithm 2 executed after the demand update of $k$. Then, we use the following properties:
    \begin{claim}
    \label{fxdT}
    It holds $f_{x,d}(T)=f(T)-x(T)$ throughout the auction.
    \end{claim}
    \begin{proof}
  By Lemma \ref{fxd_properties} (i), 
  it holds $x(T)+f_{x,d}(T)=f(T)$ at the beginning of the auction.
	Moreover, by Lemma \ref{fxd_properties} (ii),    
    the value $x(T)+f_{x,d}(T)=\min_{T'\subseteq T}\{f(T')+x(T\setminus T')+d(T\setminus T')\}$ is unchanged by Algorithm 2 (Lemma \ref{fxd_properties} (iii)) and is non-increasing during the demand updates in lines 5 and~9. This means that $x(T)+ f_{x,d}(T)$ is non-increasing as Algorithm 1 proceeds.
    At the end of the auction, by $d_i=0$ for each $i\in N$, 
    it holds $x(T)+f_{x,d}(T)=x^{\rm f}(T)$, which is equal to $f(T)$ 
    by the inductive assumption on $T$.
    Therefore, since $f_{x,d}(T)=f(T)-x(T)$ holds
    at the beginning and at the end of the auction, 
    it holds throughout the auction. 
    \end{proof}
    \begin{claim}
    \label{fx'd'T}
    It holds $f_{x',d'}(T)=f_{x',d'}(S)$.
    \end{claim}
    \begin{proof}
    By the first claim, it holds $f_{x',d'}(T)=f(T)-x'(T)$.
	Then, by Lemma~\ref{fxd_properties}~(ii) and 
    $d'_i=0$ for each $i\in T\setminus S$, it holds 
    \begin{align*}
    f_{x',d'}(S)&=\min_{S'\subseteq S}\{f(S')-x'(S')+d'(S\setminus S')\}
    =\min_{S'\subseteq S}\{f(S')-x'(S')+d'(T\setminus S')\}\\
    &\geq \min_{S'\subseteq T}\{f(S')-x'(S')+d'(T\setminus S')\}= f_{x',d'}(T).
    \end{align*}
	Combining this with the monotonicity of $f_{x',d'}$, we have $f_{x',d'}(T)=f_{x',d'}(S)$.
    \end{proof}

	By the above claims, it holds 
	$x'(S)+f_{x',d'}(S)=x'(S)+f_{x',d'}(T)=f(T)-x'(T\setminus S)$.
    Since buyers in $T\setminus S$ has already been dropped in Algorithm 2, 
    it holds $x'(T\setminus S)=x^{\rm f}(T\setminus S)$.
    Moreover, by the inductive assumption, we have $x^{\rm f}(T)=f(T)$. 
    Therefore, we have 
    \[
    x'(S)+f_{x',d'}(S)=f(T)-x'(T\setminus S)=f(T)-x^{\rm f}(T\setminus S)
    =x^{\rm f}(S).
    \]
	By Lemma \ref{fxd_properties} (ii) and (iii), 
    the value $x(S)+f_{x,d}(S)$ is unchanged 
	during the demand update of $k\notin S$ and Algorithm 2.
	Using this, we have 
	$x(S)+f_{x,d}(S)=x'(S)+f_{x',d'}(S)=x^{\rm f}(S)$.
	Therefore, we have $f_{x,d}(S)=x^{\rm f}(S)-x(S)$.

 	Secondly, we show that $f_{x,d}(S)=f(S)-x(S)$.
	Suppose to the contrary that 
        there exists a set $S'\subset S$ with 
	$f_{x,d}(S)=f(S')-x(S')+d(S\setminus S')$ 
        in Lemma \ref{fxd_properties} (ii).
	For each $j\in S\setminus S'$, it holds  
        \begin{align*}
    f_{x,d}(S\setminus j)+d_j&\leq f(S')-x(S')+d(S\setminus (S'\cup j))+d_j= f_{x,d}(S)\\
    &\leq \min_{S''\subseteq S\setminus j}\{f(S'')-x(S'')+d(S\setminus S'')\}+d_j
    =f_{x,d}(S\setminus j)+d_j,
        \end{align*}
	where the first and second inequalities hold by Lemma \ref{fxd_properties} (ii).
    Thus, $f_{x,d}(S)= f_{x,d}(S\setminus j)+d_j$.
    Moreover, by (\ref{fact}) and $k\notin S$,
	we have  
	$f_{x,d^{-k}}(S)-f_{x,d^{-k}}(S\setminus j)=f_{x,d}(S)-f_{x,d}(S\setminus j)=d_j$.
    Here we show 
    $f_{x,d^{-k}}(N)-f_{x,d^{-k}}(N\setminus j)=f_{x,d^{-k}}(S)-f_{x,d^{-k}}(S\setminus j)(=d_j),$ which implies a contradiction by $j\in S$.
    Let $l$ be a buyer in $N\setminus S$.    
    If $l\in T\setminus S$, then it holds $l\preceq k$ by the definition of $S$.
    If $l\in N\setminus T$, then $l$ is not active 
    just before the demand update of $k$ by the definition of $T$.
    In this case, we have 
	\[
 f_{x,d^{-k}}(N\setminus l)+d_l\leq f_{x,d^{-k}}(N)\leq f_{x,d^{-k}}(N\setminus l)+d_l,
 \]
	where the first inequality holds by the monotonicity of $f_{x,d^{-k}}$ and $d_l=0$, and 
	the second inequality holds by (\ref{naive}).
	From this, we have $f_{x,d^{-k}}(N)=f_{x,d^{-k}}(N\setminus l)+d_l$, 
    which means $l\preceq k$.
    Thus, in both cases, it holds $l\preceq k$.
	By iteratively applying Lemma \ref{unsaturation_property} (ii),
	it holds $f_{x,d^{-k}}(N)=f_{x,d^{-k}}(S)+d(N\setminus S)$ and 
	 $f_{x,d^{-k}}(N\setminus j)=f_{x,d^{-k}}(S\setminus j)+d(N\setminus S)$ for each $j\in S$.
	Using this, we have 
	$f_{x,d^{-k}}(N)-f_{x,d^{-k}}(N\setminus j)=f_{x,d^{-k}}(S)-f_{x,d^{-k}}(S\setminus j)=d_j$.
	Thus, it holds $j\preceq k$, contradicting the definition of $S$.
	Therefore, we have $f_{x,d}(S)=f(S)-x(S)$. 
	\end{proof}

\section{Efficiency}
In this section, we provide three types of efficiency guarantees for our mechanism.
Our tight sets lemma (Theorem \ref{tightsets}) plays critical roles in the proofs.

\subsection{Pareto Optimality}
We first show that our mechanism enjoys Pareto optimality, which has been the 
efficiency goal in many previous studies for clinching auctions with budgets. 

	\begin{theorem}
	\label{PO}
	Our mechanism satisfies PO.
	\end{theorem}

In the following, we prove Theorem \ref{PO}.
The proof is an inductive argument with respect to $\{X_k\}_{k\in [t]\cup 0}$ 
in our tight sets lemma, 
as in Goel et al. \cite{GMP2014} for their divisible setting. 
Due to the difference of the tight sets lemma, instead of dropping prices (as they used),  
we use a new non-increasing sequence $\{v^{(k)}\}_{k\in [t]}$ defined~by  
\begin{equation}
\label{theta}
v^{(k)}:=\displaystyle \min_{i\in X_{k}}v_{i}\quad (k\in [t]).
\end{equation}
Then, the following holds:  
\begin{lemma}
\label{prepare_PO}
For each $k\in[t]$, the following~holds: 
\begin{itemize}
\item[(i)] If $c^{\rm f}_{i_k}=v_{i_k}$, then it holds $c^{\rm f}_{i_k}=v^{(k)}=v_{i_k}$ 
and $B_i-p^{\rm f}_i<v^{(k)}$  for each $i\in X_k\setminus (X_{k-1}\cup i_k)$.
\item[(ii)] 
If $c^{\rm f}_{i_k}<v_{i_k}$, then it holds $c^{\rm f}_{i_k}<v^{(k)}\leq v_{i_k}$ and 
$B_i-p^{\rm f}_i<v^{(k)}$ for each $i\in X_k\setminus X_{k-1}$.
\end{itemize}
\end{lemma}

\begin{proof}
(i) 
Let $i$ be an arbitrary buyer in $X_k$ and $k'$ be a positive integer with $k'\leq k$ such that $i\in X_{k'}\setminus X_{k'-1}$.
Since $\{c^{\rm f}_{i_k}\}_{k\in [t]}$ is non-increasing, 
we have $c^{\rm f}_{i_k}\leq c^{\rm f}_{i_{k'}}=c^{\rm f}_i\leq v_i$, 
where the equality holds by Theorem \ref{tightsets} (ii) and 
the second inequality holds by the second item of Definition~\ref{dropping_prices}.
Then, if $v_{i_k}=c^{\rm f}_{i_k}(\leq v_i)$, we have $\displaystyle v^{(k)}=\min_{i\in X_k}v_i=v_{i_k}$.
Moreover, 
by Theorem~\ref{tightsets} (ii) and (iv)~$[\neg (b) \Rightarrow \neg (a)]$, we have $B_i-p_i<c^{\rm f}_{i}=c^{\rm f}_{i_k}=v^{(k)}$ for each $i\in X_k\setminus (X_{k-1}\cup i_k)$.

(ii) 
If $c^{\rm f}_{i_k}<v_{i_k}$, by Theorem \ref{tightsets} (iv) [(b) $\Rightarrow$ (c)], it holds $c^{\rm f}_{i_k}<v_{i}$ for each $i\in X_k$.
Therefore, we have $\displaystyle c^{\rm f}_{i_k}<\min_{i\in X_k}v_i\ (=v^{(k)}) \leq v_{i_k}$.
Moreover, by Theorem \ref{tightsets} (ii) and (iii), it holds $B_i-p_i\leq c^{\rm f}_{i}=c^{\rm f}_{i_k}< v^{(k)}$  
for each $i\in X_k\setminus (X_{k-1}\cup i_k)$.
By Theorem \ref{tightsets} (iv) [(b) $\Rightarrow$ (c)], 
it also holds $B_{i_k}-p_{i_k}= c^{\rm f}_{i_k}< v^{(k)}$.
\end{proof}
\begin{proof}[Proof of Theorem \ref{PO}]
Suppose to the contrary that there exists an allocation 
$\mathcal A:=(x',p')$ satisfying (i)
$v_i x^{\rm f}_i-p^{\rm f}_i\leq v_i x'_i - p'_i$ for each $i\in N$, (ii) $p^{\rm f}(N)\leq p'(N)$, and 
(iii) at least one of these inequalities holds without equality. 
Combining them, 
we have $\sum_{i\in N}v_i x^{\rm f}<\sum_{i\in N}v_i x'_i$.
In the following, we show the opposite inequality.

Let $\{v^{(k)}\}_{k\in [t]}$ be the non-increasing sequence in (\ref{theta}).
We first show that $v^{(k)} (x^{\rm f}_{i}-x'_{i}) \leq p^{\rm f}_{i}- p'_{i}$ for each $k\in [t]$ and $i\in X_{k}\setminus X_{k-1}$ by the case-by-case analysis:

Case 1: $x^{\rm f}_i\geq x'_i$. 
By $v^{(k)}=\displaystyle \min_{i'\in X_{k}}v_{i'}\leq v_i$, it holds $v^{(k)} (x^{\rm f}_i-x'_i)\leq v_i (x^{\rm f}_i-x'_i) \leq p^{\rm f}_i- p'_i$,
where the last inequality follows from assumption (i).

Case 2: $x^{\rm f}_i< x'_i$ and $c^{\rm f}_{i_k}=v_{i_k}$.
For $i=i_k$, by $c^{\rm f}_{i_k}=v_{i_k}$ and Lemma \ref{prepare_PO} (i), 
it holds $c^{\rm f}_{i_k}=v^{(k)}=v_{i_k}$.
Then, we also have 
$v^{(k)} (x^{\rm f}_{i_k}-x'_{i_k})=v_{i_k} (x^{\rm f}_{i_k}-x'_{i_k}) \leq p^{\rm f}_{i_k}- p'_{i_k}$.
For $i\neq i_k$, by Lemma \ref{prepare_PO} (i), it holds $p'_i\leq B_i<p^{\rm f}_i+v^{(k)}$. By $x^{\rm f}_i< x'_i$ and the indivisibility, 
we have $v^{(k)} (x^{\rm f}_i-x'_i)\leq -v^{(k)} < p^{\rm f}_i- p'_i$.

Case 3: $x^{\rm f}_i< x'_i$ and $c^{\rm f}_{i_k}<v_{i_k}$.
For each $i$, by Lemma \ref{prepare_PO} (ii), it holds $p'_i\leq B_i<p^{\rm f}_i+v^{(k)}$.
By $x^{\rm f}_i< x'_i$ and the indivisibility, 
we have $v^{(k)} (x^{\rm f}_i-x'_i)\leq -v^{(k)} < p^{\rm f}_i- p'_i$.

From the above, we can also see that 
for each $k\in [t]$ and $i\in X_k\setminus X_{k-1}$, 
if $v^{(k)} (x^{\rm f}_i-x'_i)=p^{\rm f}_i- p'_i$, 
then we have $v_i=v^{(k)}$ or $x^{\rm f}_i= x'_i$.

Now we show
$p^{\rm f}(X_k)-p'(X_k)\geq v^{(k)}(x^{\rm f}(X_k)-x'(X_{k}))\geq 0$
for each $k\in [t]\cup 0$ by induction, where $v^{(0)}:=v^{(1)}$.
For $k=0$, the inequality trivially holds by $X_0=\emptyset$. 
Then, we have  
\begin{align*}
0&\leq v^{(k)}(x^{\rm f}(X_{k})-x'(X_{k}))\\
&\leq v^{(k-1)}(x^{\rm f}(X_{k-1})-x'(X_{k-1}))+v^{(k)}(x^{\rm f}(X_{k}\setminus X_{k-1})-x'(X_{k}\setminus X_{k-1}))\\
&\leq p^{\rm f}(X_{k-1})-p'(X_{k-1})+v^{(k)}(x^{\rm f}(X_{k}\setminus X_{k-1})-x'(X_{k}\setminus X_{k-1}))\\
&\leq p^{\rm f}(X_{k-1})-p'(X_{k-1})+p^{\rm f}(X_{k}\setminus X_{k-1})-p'(X_{k}\setminus X_{k-1}))\\
&=p^{\rm f}(X_{k})-p'(X_{k}),
\end{align*}
where the first inequality follows from $x'(X_k)\leq f(X_k)=x^{\rm f}(X_{k})$ (Theorem \ref{tightsets} (i)), 
the second inequality follows from $v^{(k)}\leq v^{(k-1)}$ and $x'(X_{k-1})\leq f(X_{k-1})=x^{\rm f}(X_{k-1})$, 
the third inequality follows by induction, and the fourth inequality follows by 
$v^{(k)} (x^{\rm f}_{i}-x'_{i}) \leq p^{\rm f}_{i}- p'_{i}$ for each 
$k\in [t]$ and $i\in X_{k}\setminus X_{k-1}$.
By substituting $k$ with $t$, we have $p^{\rm f}(N)-p'(N)\geq v^{(t)}(x^{\rm f}(N)-x'(N))\geq 0.$
Since we assume $p^{\rm f}(N)\leq p'(N)$, all the inequalities hold in equality. Therefore, 
we have $v_i=v^{(k)}$ or $x^{\rm f}_i= x'_i$ 
for each $k\in [t]$ and $i\in X_k\setminus X_{k-1}$.

Using this, we show $\sum_{i\in X_{k}}v_i (x^{\rm f}_i-x'_i)\geq v^{(k)}(x^{\rm f}(X_{k})-x'(X_{k}))\geq 0$  for each $k\in [t]$ by induction.
For $k=0$, the inequality trivially holds by $X_0=\emptyset$.  
Then, we have
\begin{align*}
\sum_{i\in X_{k}}v_i (x^{\rm f}_i-x'_i)&\geq v^{(k-1)} (x^{\rm f}(X_{k-1})-x'(X_{k-1}))+\sum_{i\in X_{k}\setminus X_{k-1}}v^{(k)} (x^{\rm f}_i-x'_i)\\
&\geq v^{(k)} (x^{\rm f}(X_{k-1})-x'(X_{k-1}))+\sum_{i\in X_{k}\setminus X_{k-1}}v^{(k)} (x^{\rm f}_i-x'_i)\\
&= v^{(k)} (x^{\rm f}(X_{k})-x'(X_{k}))\geq 0,
\end{align*}
where the first inequality holds by induction, and $v_i=v^{(k)}$ or $x^{\rm f}_i= x'_i$ for each $i\in X_{k}\setminus X_{k-1}$, 
the second inequality holds by the non-increasing of $v^{k}$ 
and the tightness of $X_{k-1}$ (Theorem \ref{tightsets}~(i)), 
and the last inequality holds by the tightness of $X_{k}$.
Thus, it holds $\sum_{i\in X_{k}}v_i (x^{\rm f}_i-x'_i)\geq v^{(k)} (x^{\rm f}(X_{k})-x'(X_{k}))\geq 0$ for any $k\in [t]\cup 0$.
Substitute $k$ with $t$, we have $\sum_{i\in N}v_i x^{\rm f}_i\geq \sum_{i\in N}v_i x'_i$, which contradicts the hypothesis. 
\end{proof}

\begin{remark}
In Theorem \ref{PO}, we used the tight sets lemma to show PO. 
This is different from the previous approach of indivisible settings (Fiat et al. \cite{FLSS2011} and Colini-Baldeschi et al. \cite{BHLS2015}). They characterized and proved PO by means of the {\it non-trading path property}. In Appendix A, we show that this property can be generalized to polymatroid environments and that the generalized one can be derived from our tight sets lemma. This means that our tight sets lemma is a stronger basis for efficiency.
\end{remark}

\subsection{Liquid Welfare and Social Welfare}
	Here we establish LW and SW guarantees for clinching auctions of indivisible goods. 
        Let ${\rm LW}^{\rm M}$ and ${\rm SW}^{\rm M}$ denote the LW and SW of 
	our mechanism, respectively.
        Let ${\rm LW}^{\rm OPT}$ denote the optimal LW value, i.e., the optimal value of the following problem:
	\begin{align}
	\label{optimization}
	&{\text{maximize}}\ \ \ \ \sum_{i\in N}\min(v_i x_i,B_i) \ \ \ \ \text{subject to}\ \  x\in P\cap \mathbb Z^N.
	\end{align}
 
	Our mechanism achieves 2-approximation to the optimal LW value.
	\begin{theorem}
	\label{LW}
        It holds ${\rm LW}^{\rm M}\geq \frac{1}{2}{\rm LW}^{\rm OPT}$.
	\end{theorem}
 
    Although our LW guarantee is tight, there is a gap between our guarantee and a lower bound.
    This can be seen from the following proposition:

    \begin{proposition}
    \label{example}
    The following holds:
    \begin{itemize}
    \item[(i)] 
    There is an instance that satisfies ${\rm LW}^{\rm M}= \frac{k}{2k-1}{\rm LW}^{\rm OPT}$, where $k$ is the number of total goods.
    \item[(ii)] 
    There is no mechanism ${\rm M}'$ that satisfies all of IC, IR, and ${\rm LW}^{{\rm M}'}\geq \frac{3}{4}{\rm LW}^{\rm OPT}$, where ${\rm LW}^{{\rm M}'}$ denotes the LW of the mechanism ${\rm M}'$.
    \end{itemize}
    \end{proposition}
 	\begin{proof}
    (i) Consider a market where 
	the seller owns $k$\ $(k\gg 1)$ units of the good and 
	two buyers participate in the auction.
	Buyer 1 has valuation $1$, and buyer 2 has valuation $k$ 
	and the budgets of both buyers are $k$.
    The integer polymatroid $P$ is defined by 
    $P:=\{x\in\mathbb Z^2_{+}\mid x_1+x_2\leq k\}$.
	Then, LW is maximized when $(x_1,x_2)=(k-1, 1)$.
	Thus, the optimal LW value is $2k-1$.
	In Algorithm 1, the price increases to 1 in line 3 of the first iteration, 
	and the demand of buyer 1 decreases to zero.
	Then, all the goods are allocated to buyer 2, and 
	thus the LW of our mechanism is $\min(k^2, k)=k$.
	Therefore, 
	${\rm LW}^{\rm M}=k=\frac{k}{2k-1}{\rm LW}^{\rm OPT}$.

	(ii) Consider the same market as (i), where the valuations of buyers can be changed.
	Suppose that there exists a mechanism ${\rm M}'$ satisfying IC and IR and achieving 
	$\gamma$-approximation to the optimal LW value. 
	Let $\bar{x}(v_1,v_2)$ denote the allocation of goods by ${\rm M}'$ in the market where the valuations of buyer 1 and buyer 2 are $v_1$ and $v_2$, respectively.
	
	Since the mechanism satisfies IC, 
        by Myerson's lemma \cite{M1981}, 
        $\bar{x}_i(v_1,v_2)$ must be non-decreasing in $v_i$ for each buyer $i$.
        Then, we have
	 $\bar{x}_1(1, k)\leq \bar{x}_1(k, k)$ and  $\bar{x}_2(k, 1)\leq \bar{x}_2(k, k)$.
	 Thus, by $\bar{x}_1(k, k)+\bar{x}_2(k, k)\leq k$ and symmetry, we can assume that 
	 $\bar{x}_1(1, k)\leq \bar{x}_1(k, k)\leq \lfloor \frac{k}{2}\rfloor$ without loss of generality.
      When $(v_1,v_2)=(1,k)$, the optimal LW value is $2k-1$ as in (i).
	 Then, the LW of the allocation $\bar{x}(1, k)$ is at most 
	 \[
        \min( \bar{x}_1(1, k), k)+\min( k \bar{x}_2(1, k), k)\leq \Bigl\lfloor \frac{k}{2}\Bigr\rfloor + k \leq 
	\frac{3k}{2}=\frac{3}{4-\frac{2}{k}}(2k-1),
	 \]
	 which implies $\gamma\leq \frac{4}{3}$ by taking $k$ to $\infty$.
	\end{proof}
     Note that Proposition \ref{example} (ii) is a natural extension of Theorem 5.1 of Dobzinski and Leme \cite{DL2014} to the indivisible setting. 

    We also show that our mechanism achieves SW more than the optimal LW value.

 	\begin{theorem}
	\label{SW}
    It holds ${\rm SW}^{\rm M}\geq {\rm LW}^{\rm OPT}$.
	\end{theorem}
 
    This guarantee is tight and achieves the best possible 
    because the optimal LW value is equal to the optimal SW value  
    if the budgets of all buyers are sufficiently large. 
    In such a case, our mechanism outputs an allocation that maximizes the SW.

\subsubsection{Proof Outline}
    To prove Theorems \ref{LW} and \ref{SW}, 
    we follow the outline of the one in Sato \cite{S2023}: 
    In Theorem \ref{LW}, we show the following relationship.
    Obviously, Theorem \ref{LW2} implies Theorem \ref{LW}.  
    \begin{theorem}
	\label{LW2}
        It holds ${\rm LW}^{\rm M}\geq p^{\rm f}(N)\geq {\rm LW}^{\rm OPT}-{\rm LW}^{\rm M}$.
	\end{theorem}

	We first establish an explicit formula of an optimal LW allocation. In our indivisible setting, the formula is more complicated than that in Sato \cite{S2023}. 
    This is due to the following situation that never happens in his divisible setting: If $x_i= \left\lfloor B_i/v_i \right\rfloor$\ $(i\in N)$ and an additional unit is allocated to~$i$, the increase of $\min(v_i x_i, B_i)$ is $B_i-\left\lfloor  B_i/v_i \right\rfloor v_i$.
For such an increase, we set up another buyer who receives only the $\left\lfloor B_i/v_i \right\rfloor+1$-th unit.
Specifically, divide each buyer $i \in N$ into two {\it virtual} buyers $i_a, i_b$. 
Their valuations and budgets are set~as
\begin{align*}
& v_{i_a} := v_i, \quad   B_{i_a} := \left\lfloor  B_i/v_i \right\rfloor v_i, \\
& v_{i_b} = B_{i_b} := B_i-\left\lfloor  B_i/v_i \right\rfloor v_i.
\end{align*}
To make $v_{i_b}$ positive, 
if $B_i-\left\lfloor  B_i/v_i \right\rfloor v_i = 0$, then set $v_{i_b}$ by an arbitrary positive value $< v_i$.
Then, it holds $B_{i_a}/v_{i_a}=\lfloor B_{i}/v_{i} \rfloor$, 
$B_{i_b}/v_{i_b}=1$ if $B_i-\left\lfloor  B_i/v_i \right\rfloor v_i \neq 0$, and $B_{i_b}/v_{i_b}=0$ otherwise.

Let $N':= \cup_{i\in N}\{i_a, i_b\}$ denote the set of all virtual buyers.
Define a map $\Gamma:2^{N'}\to 2^N$ 
by $\Gamma(S') := \{i\in N \mid S' \cap \{i_a,i_b\}\neq \emptyset\}$, and
define a new monotone submodular function $f': 2^{N'}\to \mathbb Z_{+}$ by 
$f'(S'):=f(\Gamma(S'))$.  
Note that $f'$ is again a monotone submodular function; see, e.g., Section~44.6g of Schrijver \cite{S2003}.

Then, an LW-optimal allocation for (\ref{optimization}) can be obtained by the greedy procedure described as follows.
 Suppose that buyers in $N'$ are ordered and numbered in descending order of their valuations, 
 i.e., $N' = \{1,2,\ldots,2|N|\}$ and $v_{i'} \geq v_{j'}$ for $i' \leq j'$. Then, the following holds:
	\begin{proposition}
	\label{optimal}
	An LW-optimal allocation $x^*$ is given by
        \begin{align}
	x^*_i&=z^{*}_{i_{a}}+z^*_{i_{b}}\quad (i\in N),  \nonumber\\ 
        \label{optimal2}
	z^{*}_{i'}&=\min\Bigl( B_{i'}/v_{i'} ,{\min_{H\subseteq [i'-1]}\{f'(H\cup i')-z^*(H)}\}\Bigr) \qquad (i'=1,2,\ldots,N').
        \end{align}
	\end{proposition}
Note that $B_{i'}/v_{i'}$ is an integer by the construction of virtual buyers.
	Now we prove Theorem \ref{LW2} (Theorem \ref{LW}) and Theorem \ref{SW} using the following properties.

    \begin{lemma}
    \label{prepare}
    Let $x^*$ and $z^*$ be the vectors in Proposition \ref{optimal}. 
    For each $i\in N$, the following holds:
    \begin{itemize}
    \item[(i)]
	 It holds $\min(v_i x^*_i, B_i)=v_{i_a} z^*_{i_a}+v_{i_b} z^*_{i_b}$.
    \item[(ii)]
	 It holds 
	$v_i x^{\rm f}_i-(v_{i_{a}} z^{*}_{i_{a}}+v_{i_{b}} z^*_{i_{b}})\geq c^{\rm f}_i(x^{\rm f}_i-x^*_i)$.
    \item[(iii)] If $x^{\rm f}_i< x^*_i$, then it holds 
	$c^{\rm f}_{i}z_{i_{a}}^*+v_{i_{b}}z_{i_{b}}^*-c^{\rm f}_{i}x^{\rm f}_{i}= \min(v_i x^*_i, B_i)-\min(v_i x^{\rm f}_i, B_i)$.
    \end{itemize} 
	\end{lemma}

	\begin{lemma}
	\label{payment}
	It holds
	$\displaystyle \sum_{i\in N: x^{\rm f}_i\geq x^*_i}p^{\rm f}_i\geq \sum_{i\in N: x^{\rm f}_i< x^*_i}
	(c^{\rm f}_{i}z_{i_{a}}^*+v_{i_{b}}z_{i_{b}}^*-c^{\rm f}_{i}x^{\rm f}_{i}).$
	\end{lemma}

\begin{proof}[Proof of Theorem \ref{LW2}]
	The first inequality is easy by  
	\begin{equation}
	\label{lowerbound}
	{\rm LW}^{\rm M}=\sum_{i\in N}\min(v_i x^{\rm f}_i, B_i)\geq p^{\rm f}(N), 
	\end{equation}
	where 
	$v_i x^{\rm f}_i \geq p^{\rm f}_i$ holds by IR of buyers (Theorem \ref{IC_IR}) and 
	$B_i\geq p^{\rm f}_i$ holds by budget feasibility (Theorem \ref{BF}) for each $i\in N$.
    For the second inequality, we have 
	\begin{align*}
	p^{\rm f}(N) &\geq \sum_{i\in N: x^{\rm f}_i\geq x^*_i}p^{\rm f}_i\geq
 \sum_{i\in N: x^{\rm f}_i< x^*_i}(c^{\rm f}_{i}z_{i_{a}}^*+v_{i_{b}}z_{i_{b}}^*-c^{\rm f}_{i}x^{\rm f}_{i})=\sum_{i\in N: x^{\rm f}_i< x^*_i}\bigl(\min(v_i x^*_i, B_i)-\min(v_i x^{\rm f}_i, B_i)\bigr)\\
	&\geq  \sum_{i\in N}\bigl(\min(v_i x^*_i, B_i)-\min(v_i x^{\rm f}_i, B_i)\bigr)={\rm LW}^{\rm OPT}-{\rm LW}^{\rm M}, 
	\end{align*}
	where the first inequality holds by $p^{\rm f}_i\geq 0\ (i\in N)$, the first equality (resp. the second inequality) holds by Lemma~\ref{prepare} (iii) (resp. Lemma~\ref{payment}),  
     and the third inequality holds by $\min(v_i x^*_i, B_i)-\min(v_i x^{\rm f}_i, B_i)\leq 0$ for each 
    $i\in N$ with $x^{\rm f}_i\geq x^*_i$. 
    Therefore, using (\ref{lowerbound}), we have 
    ${\rm LW}^{\rm M}\geq p^{\rm f}(N)\geq {\rm LW}^{\rm OPT}-{\rm LW}^{\rm M}$.
	\end{proof}

\begin{proof}[Proof of Theorem \ref{SW}]
    By Lemma~\ref{prepare} (i) and (ii), we have
    \begin{equation}
    \label{SW-LW}
	{\rm SW}^{\rm M}- {\rm LW}^{\rm OPT}
	=\sum_{i\in N}\bigl(v_i x^{\rm f}_i-(v_{i_{a}} z^{*}_{i_{a}}+v_{i_{b}} z^*_{i_{b}})\bigr)
    \geq\sum_{i\in N}c^{\rm f}_i(x^{\rm f}_i-x^*_i).
	\end{equation}

    Now we prove 
     $\sum_{i\in X_k}c^{\rm f}_i(x^{\rm f}_i-x^*_i)
     \geq c^{\rm f}_{i_k} (x^{\rm f}(X_k)-x^*(X_k))\geq 0$
	for each $k\in [t]\cup 0$ by induction (where we set $c^{\rm f}_{i_0}:=c^{\rm f}_{i_1}$). 
	For $k=0$, the above inequality holds by $X_0=\emptyset$.
	By Theorem~\ref{tightsets} (ii), we have 
	$c^{\rm f}_{i}(x^{\rm f}_i-x^*_i)= c^{\rm f}_{i_k}(x^{\rm f}_i-x^*_i)$ 
	for $i\in X_{k}\setminus X_{k-1}$.
	Therefore, we have
	\begin{align*}
	\sum_{i\in X_{k}}c^{\rm f}_{i}(x^{\rm f}_i- x^*_i)
    &\geq c^{\rm f}_{i_{k-1}} (x^{\rm f}(X_{k-1})-x^*(X_{k-1}))+
	 c^{\rm f}_{i_k}\sum_{i\in X_{k}\setminus X_{k-1}}(x^{\rm f}_i- x^*_i) \\
	&\geq c^{\rm f}_{i_{k}} (x^{\rm f}(X_{k})-x^*(X_{k}))\geq 0, 
	\end{align*}
	where the first inequality holds by induction, 
    the second inequality holds by $c^{\rm f}_{i_{k-1}}\geq c^{\rm f}_{i_{k}}$ 
    and $x^*(X_{k-1})\leq f(X_{k-1})=x^{\rm f}(X_{k-1})$ (Theorem \ref{tightsets} (i)), 
    and the last inequality holds by $x^*(X_{k})\leq f(X_{k})=x^{\rm f}(X_{k})$.
	Substituting $k$ with $t$, by $X_t=N$, we have 
    $\sum_{i\in N}c^{\rm f}_{i}(x^{\rm f}_i- x^*_i)\geq 0.$
    Therefore, using (\ref{SW-LW}), we have 
    ${\rm SW}^{\rm M}-{\rm LW}^{\rm OPT}\geq \sum_{i\in N}c^{\rm f}_{i}(x^{\rm f}_i- x^*_i)\geq 0.$ 
    \end{proof}
    Note that in the proof of Theorem \ref{SW}, 
    we use $\{c^{\rm f}_{i_k}\}$ instead of 
    $\{v_{i_k}\}$ used in Sato \cite{S2023} because 
    $\{v_{i_k}\}$ is not necessarily monotone 
    in our indivisible setting.
     
\begin{figure}[ht]
\centering
\begin{tikzpicture}[
  node distance=1.25cm and 2cm,
  every node/.style={draw, rectangle, minimum height=0.75cm, minimum width=2.5cm, align=center,font=\small},
  -{Latex[length=3mm]} 
]

\node (thm42) {Theorem 4.2\\ (Tight Sets Lemma)};

\node (lemma512) [below left=of thm42] {Lemma 5.12\\ (Remaining Goods)};
\node (lemma511) [below=of thm42] {Lemma 5.11};
\node (lemma510) [below right=of thm42] {Lemma 5.10};

\node (lemma59) [below=of lemma512] {Lemma 5.9\\ (Total Payments)};
\node (lemma58) [below=of lemma511] {Lemma 5.8};
\node (prop57) [below=of lemma510] {Proposition 5.7\\ (LW Optimal Allocation)};

\node (thm53) [below=of $(lemma59)!0.5!(lemma58)$] {Theorem 5.3\\ (LW Guarantee)}; 
\node (thm55) [below=of $(lemma58)!0.5!(prop57)$] {Theorem 5.5\\ (SW Guarantee)}; 

\draw[-{Triangle[length=1.5mm, width=1.5mm]}] (thm42) -- (lemma511);
\draw[-{Triangle[length=1.5mm, width=1.5mm]}]  (lemma510) -- (lemma511);
\draw[-{Triangle[length=1.5mm, width=1.5mm]}]  (lemma510) -- (prop57);
\draw[-{Triangle[length=1.5mm, width=1.5mm]}]  (prop57) -- (lemma58);
\draw[-{Triangle[length=1.5mm, width=1.5mm]}]  (prop57) -- (lemma511);
\draw[-{Triangle[length=1.5mm, width=1.5mm]}]  (lemma510) -- (lemma58);
\draw[-{Triangle[length=1.5mm, width=1.5mm]}]  (lemma511) -- (lemma58);
\draw[-{Triangle[length=1.5mm, width=1.5mm]}]  (lemma511) -- (lemma59);
\draw[-{Triangle[length=1.5mm, width=1.5mm]}]  (lemma512) -- (lemma59);
\draw[-{Triangle[length=1.5mm, width=1.5mm]}]  (lemma59) -- (thm53);  
\draw[-{Triangle[length=1.5mm, width=1.5mm]}]  (lemma58) -- (thm53);  
\draw[-{Triangle[length=1.5mm, width=1.5mm]}]  (lemma58) -- (thm55);  
\draw[-{Triangle[length=1.5mm, width=1.5mm]}]  (thm42) -| ($(thm42)+(7,0)$) |- (thm55); 
\end{tikzpicture}
\caption{Diagram of dependencies between theorems, propositions, and lemmas in Section 5.2.}
\label{outline_proof}
\end{figure}
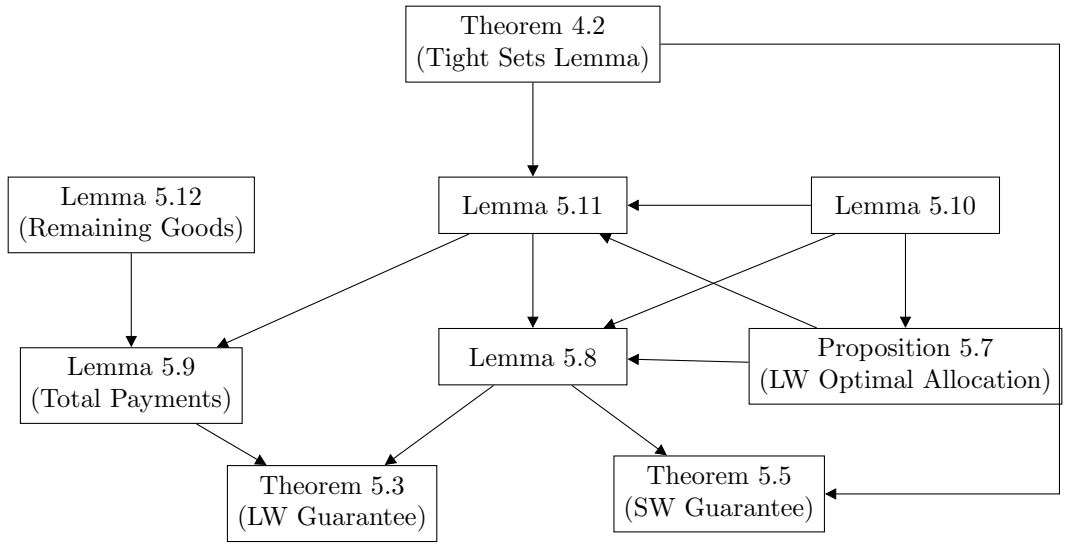

The remainder of this section is devoted to proving 
Proposition \ref{optimal} and Lemmas \ref{prepare} and \ref{payment},
where the outline of the proofs is illustrated in Figure \ref{outline_proof}.
In Section 5.2.2, we show Proposition \ref{optimal}.
In Section 5.2.3, we show Lemma \ref{prepare} 
using the tight sets lemma (Theorem \ref{tightsets}).
In Section 5.2.4, we show Lemma \ref{payment} 
via lower bounding the number of remaining goods and the future payments at each moment of Algorithm 1.

\subsubsection{Proof of Proposition \ref{optimal}}
        We first show that the formula (\ref{optimal2}) provides an LW-optimal allocation 
        for virtual buyers. Consider the following LW maximizing problem for virtual buyers:
	\begin{align}
	\label{opt2}
	{\text{maximize}}\ \ \ \ \sum_{i'\in N'}v_{i'} \min(z_{i'},  B_{i'}/v_{i'} )\quad
	\text{subject to}\ \ \ \, z\in P(f')\cap \mathbb Z^{N'}.
	\end{align}
	Indeed, we can find an optimal solution by solving 
	\begin{align}
	\label{opt1}
	{\text{maximize}}\ \ \ \ \sum_{i'\in N'}v_{i'} z_{i'}\quad
	\text{subject to}\ \ \ \, z\in P_{d'}(f')\cap \mathbb Z^{N'},\qquad\qquad \nonumber
	\end{align}
	where $d':=(d'_{i'})_{i'\in N'}$ with $d'_{i'}:=B_{i'}/v_{i'}$ and 
	$P_{d'}(f'):=\{y\in P(f')\mid y_{i'}\leq d'_{i'}\ (i'\in N')\}$.
	Note that $P_{d'}(f')$ is a reduction of an integer polymatroid $P$ by an integer vector 
	$d'$ and is again an integer polymatroid; see Section 3.1 of Fujishige \cite{F2005}. 
        The monotone submodular function $f'_{d'}:2^{N'}\to \mathbb Z_{+}$ that defines $P_{d'}(f')$ is given by
        $f'_{d'}(S)=\min_{S'\subseteq S}\{f'(S\setminus S')+d'(S')\}\ (S\subseteq N')$.
	Therefore, this problem is a linear optimization on polymatroids and 
	is known to be solved efficiently by a greedy procedure: An optimal solution 
    $z^{*}:=(z^*_i)_{i'\in N'}$ of (\ref{opt2}) is obtained by $z^*_{i'}:=f'_{d'}([i']) - z^*([i'-1])\ \ (i' = 1,2,\ldots, N')$.
	Moreover, by the integrality of polymatroids, 
        we can apply the same arguments of Proposition 2.1 in Sato \cite{S2023}.
        This means that $z^{*}$ is also expressed by  (\ref{optimal2}).\footnote{Using his arguments, we can show that (i) if $f'_{d'}([i'])=f'_{d'}([i'-1])+d'_{i'}$, then it holds $z^*_{i'}=d'_{i'}$, and (ii) if $f'_{d'}([i'])<f'_{d'}([i'-1])+d'_{i'}$, then $z^*_{i'}$ is the minimum of $f'(H\cup i')-z^*(H)$ with respect to $H\subseteq [i'-1]$. Then, (\ref{optimal2}) is obtained by taking the minimum of both cases.}

        It remains to show that $x^*_i=z^*_{i_a}+z^*_{i_b}\ (i\in N)$ provides an optimal solution of the LW maximizing problem (\ref{optimization}) for original buyers.
        We use the following lemma:
	\begin{lemma}
	\label{prepare_optimal}
	The following holds: 
        \begin{itemize}
        \item[(i)]  For $x_i\in \mathbb Z_+$, it holds 
        \[\min(v_i x_i,B_i)=v_{i}\min(x_{i},\left\lfloor  B_i/v_i \right\rfloor)+\bigl(B_i-v_i\left\lfloor  B_i/v_i \right\rfloor\bigr)\min\bigl(\max(x_{i}-\left\lfloor  B_i/v_i \right\rfloor,0),B_{i_b}/v_{i_b}\bigr).\]
        \item[(ii)] 
        If $z^*_{i_{b}}= B_{i_b}/v_{i_b} =1$, then $z^{*}_{i_{a}}= B_{i_a}/v_{i_a} =\left\lfloor  B_i/v_i \right\rfloor$. 
        \item[(iii)] It holds $\min(v_i x^*_i, B_i)=v_{i_a} z^*_{i_a}+v_{i_b} z^*_{i_b}$.
	\end{itemize}
        \end{lemma} 
 	\begin{proof}
        (i) If $B_i-v_i\left\lfloor  B_i/v_i \right\rfloor=0$, i.e., $B_i/v_i=\left\lfloor  B_i/v_i \right\rfloor$, then the equation trivially holds. Otherwise, it holds 
        $B_{i_b}/v_{i_b}=1$. 
        If $v_i x_i\leq B_i$, by $x\in\mathbb Z^N_+$, 
        it holds $x_i\leq\left\lfloor  B_i/v_i \right\rfloor$, 
        and then both sides are equal to $v_i x_i$.
        If $v_i x_i> B_i$, it holds $x_i\geq\left\lfloor  B_i/v_i \right\rfloor+1$.
        Then, the right-hand side is equal to 
        $v_{i}\left\lfloor  B_i/v_i \right\rfloor+(B_i-v_i\left\lfloor  B_i/v_i \right\rfloor)=B_i=\min(v_i x_i,B_i)$.

	(ii) By (\ref{optimal2}), it holds 
 $z^{*}_{i_{a}}\leq\left\lfloor B_i/v_i \right\rfloor$ and $z^*_{i_{b}}\in \{0, 1\}$.
We show the contraposition:  If $z^{*}_{i_{a}}<\left\lfloor  B_i/v_i \right\rfloor$, then $z^*_{i_{b}}=0$.
	Suppose that $z^{*}_{i_{a}}<\left\lfloor  B_i/v_i \right\rfloor$. Then, by (\ref{optimal2}), there exists  $H\subseteq [i_{a}-1]$ such that 
	$z^*(H\cup i_{a})=f'(H\cup i_{a})$.
	By the definition of $f'$ and $\Gamma$, 
	it holds $f'(H\cup i_{a}\cup i_{b})=
        f(\Gamma(H)\cup i)
=f(\Gamma(H\cup i_{a})) =f'(H\cup i_{a})$.
        Moreover, by $v_{i_{a}}=v_i>v_{i_{b}}$ and the ordering of $N'$, 
        we have $[i_{a}-1]\cup i_{a} \subseteq [i_{b}-1]$. 
        Therefore, by (\ref{optimal2}), we have 
	$z^*_{i_{b}}\leq\min_{H\subseteq [i_{b}-1]}\{f'(H\cup i_{b})-z^*(H)\}
	\leq f'(H\cup i_{a} \cup i_{b})-z^*(H\cup i_{a})=f'(H\cup i_{a})-z^*(H\cup i_{a})=0$.

        (iii) If $v_i x^*_i\leq B_i$, it holds $x^*_i\leq B_i/v_i < \left\lfloor  B_i/v_i \right\rfloor+1
        =B_{i_a}/v_{i_a}+B_{i_b}/v_{i_b}$. Then, by (ii) and (\ref{optimal2}), we have $z^{*}_{i_b}=0$. Therefore, by $v_i=v_{i_a}$, 
        it holds $\min(v_i x^*_i, B_i)=v_{i_a}z^{*}_{i_a}=v_{i_a}z_{i_{a}}^*+v_{i_{b}}z_{i_{b}}^*$.
        If $v_i x^*_i> B_i$, it holds $x^*_i> B_i/v_i \geq \left\lfloor  B_i/v_i \right\rfloor$. Also, by (\ref{optimal2}), it holds 
        $x^*_i=z^{*}_{i_a}+z^{*}_{i_b} \leq \left\lfloor  B_i/v_i \right\rfloor+1$.
        Therefore, by $x^*_i\in \mathbb Z_+$, 
        it holds 
        $x^*_i=\left\lfloor  B_i/v_i \right\rfloor+1$.
        By (\ref{optimal2}), this implies 
        $z^{*}_{i_a}=\left\lfloor  B_i/v_i \right\rfloor$ and $z^{*}_{i_b}=1$.
        By the definition of $v_{i_a}$ and $v_{i_b}$, we have $\min(v_i x^*_i, B_i)=B_i
        v_{i}\left\lfloor  B_i/v_i \right\rfloor+(B_i-v_{i}\left\lfloor  B_i/v_i \right\rfloor)
        =v_{i_a}z^{*}_{i_a}+v_{i_b}z^{*}_{i_b}$.
	\end{proof}

\begin{proof}[Proof of Proposition \ref{optimal}]
        We show that an optimal allocation of (\ref{opt2}) can be transformed into a feasible allocation of (\ref{optimization}) without changing the objective value, and vice versa.
        This leads to the conclusion that the optimal values of both problems are equal.

        Let $z^{*}\in P(f')\cap \mathbb Z^{N'}_{+}$ be an optimal allocation of 
        (\ref{opt2}) obtained by (\ref{optimal2}).
	Define $x^*:=(x^*_i)_{i\in N}$ with $x^*_i=z^*_{i_a}+z^*_{i_b}$ for each $i\in N$.
        Then, for each $S\subseteq N$, we have $x^*(S)=\sum_{i\in S}(z^{*}_{i_a}+z^{*}_{i_b})=z^{*}(\cup_{i\in S}\{i_a,i_b\})\leq f'(\cup_{i\in S}\{i_a,i_b\})=f(S)$, 
        where the inequality holds by $z^{*}\in P(f')$ and the last equality holds by the definition of $f'$ and $\Gamma$. 
        From this, we have $x^*\in P(f)\cap \mathbb Z^N_{+}$.
        Therefore, $x^*$ is a feasible allocation of (\ref{optimization}).
        For the objective value, we have 
 	\begin{align*}
        \sum_{i'\in N'}v_{i'}\min\bigl(z^*_{i'},  B_{i'}/v_{i'} \bigr)
	=\sum_{i'\in N'}v_{i'}z^*_{i'}=\sum_{i\in N}(v_{i_a}z^{*}_{i_a}+v_{i_b}z^{*}_{i_b})=
        \sum_{i\in N}\min(v_i x^*_i,B_i),
	\end{align*}
 	where the first equality holds by (\ref{optimal2}) and the third equality holds 
        by Lemma~\ref{prepare_optimal} (iii). This means that after the transformation to $x^*$, 
        the objective value is unchanged.
        
	For a vector $x\in  P(f)\cap \mathbb Z^N_{+}$, define 
	$z_{i_a}=\min(x_i, \left\lfloor  B_i/v_i \right\rfloor)$ and 
        $z_{i_b}=x_i-z_{i_a}=\max(x_i-\left\lfloor  B_i/v_i \right\rfloor,0)$ 
        for each $i\in N$. By construction, it holds $z\in \mathbb Z^{N'}_{+}$ and 
        $z_{i_a}+z_{i_b}=x_i$ for each $i\in N$. 
	Then, by $x\in P(f)\cap \mathbb Z^N_+$ and the definition of $f'$ and $\Gamma$, 
        we have $z(S')\leq x_i(\Gamma (S'))
        \leq f(\Gamma (S'))=f'(S')$ for each $S'\subseteq N'$,
        which implies $z\in P(f') \cap \mathbb Z^{N'}_+$.
        Moreover, we have 
	\begin{align*}
	\sum_{i\in N}\min(v_i x_i,B_i)&=\sum_{i\in N}\Bigl(v_{i}\min(x_i,\left\lfloor  B_i/v_i \right\rfloor)+\bigl(B_i-v_i\left\lfloor  B_i/v_i \right\rfloor\bigr)\min\bigl(\max(x_i-\left\lfloor  B_i/v_i \right\rfloor,0),B_{i_b}/v_{i_b}\bigr)\Bigr)\\
	&=\sum_{i\in N}\Bigl(v_{i_a}\min\bigl(x_i,\left\lfloor  B_i/v_i \right\rfloor\bigr)+v_{i_b}\min(\max(x_i-\left\lfloor  B_i/v_i \right\rfloor,0),B_{i_b}/v_{i_b})\Bigr)\\
 	&=\sum_{i\in N}\Bigl(v_{i_a}\min\bigl(z_{i_a},\left\lfloor  B_i/v_i \right\rfloor\bigr)+v_{i_b}\min(z_{i_b},B_{i_b}/v_{i_b})\Bigr)\\
   	&=\sum_{i\in N}\Bigl(v_{i_a}\min\bigl(z_{i_a},  B_{i_a}/v_{i_a} \bigr)+v_{i_b}\min(z_{i_b},
     B_{i_b}/v_{i_b} )\Bigr)
       =\sum_{i'\in N'}v_{i'}\min\bigl(z_{i'},  B_{i'}/v_{i'} \bigr),
	\end{align*}
        where the first equality holds by Lemma~\ref{prepare_optimal} (i), 
        the second and fourth equalities hold by the construction of virtual buyers, and  
        the third equality holds by the construction of $z_{i_a}$ and $z_{i_b}$.
        Therefore, an optimal solution of (\ref{optimization}) can also be transformed into 
        a feasible solution of (\ref{opt2}) without changing the objective value.

        By the above, we can see that the optimal values of both problems are equal.	
        Therefore, calculating $z^{*}$ by (\ref{optimal2}) and defining $x^*_i=z^*_{i_a}+z^*_{i_b}\ (i\in N)$ lead to an optimal allocation of~(\ref{optimization}).
\end{proof}

\subsubsection{Proof of Lemma \ref{prepare}}
    In the proof, we use the following lemma, which illustrates the relationship between the allocation of goods in our mechanism and the optimal allocation (\ref{optimal2}).

	\begin{lemma}
	\label{algo<opt}
	If $x^{\rm f}_i<x^*_i$ and $c^{\rm f}_i<v_i$, then it holds $x^{\rm f}_i=z^{*}_{i_{a}}=\left\lfloor  B_i/v_i \right\rfloor$,
	$z^*_{i_{b}}=1$, and $v_{i_{b}}\leq c^{\rm f}_i$.
	\end{lemma}
    We will first prove this lemma and then give the proof of Lemma \ref{prepare}.
	\begin{proof}
        First note that, in Algorithm~1, buyers clinch their allocated goods at a price less than or equal to 
        their dropping prices. Then, it holds 
 	\begin{align}
	\label{x}	
     x^{\rm f}_i\geq 
     \left\lceil p^{\rm f}_i/c^{\rm f}_i\right\rceil\geq 
     \left\lceil p^{\rm f}_i/v_i\right\rceil
     \quad (i\in N), 
     \end{align}
     where the last inequality holds by $c^{\rm f}_i\leq v_i$ (Definition \ref{dropping_prices}).
    
	 Suppose that $x_i^{\rm f}<x_i^*$ and $c_i^{\rm f}<v_i$. 
    By $c^{\rm f}_i<v_i$ and Theorem~\ref{tightsets}~(iv) [(b)~$\Rightarrow$~(c)], it holds $(B_i-p^{\rm f}_i)/c^{\rm f}_i\leq 1$
    for $i=i_k$. For $i\neq i_k$,
    by Theorem \ref{tightsets} (iii), it also holds $(B_i-p^{\rm f}_i)/c^{\rm f}_i\leq 1$.
    Then, again by $c^{\rm f}_i<v_i$, we have 
    $\left\lfloor (B_i-p^{\rm f}_i)/v_i \right\rfloor\leq 
	(B_i-p^{\rm f}_i)/v_i<(B_i-p^{\rm f}_i)/c^{\rm f}_i\leq 1$.
    Thus, we have $\left\lfloor (B_i-p^{\rm f}_i)/v_i\right\rfloor =0$.
	Using this and (\ref{x}), we have 
	$x^{\rm f}_i\geq \left\lceil p^{\rm f}_i/v_i\right\rceil \geq \left\lfloor  B_i/v_i \right\rfloor=B_{i_a}/v_{i_a}\geq z^{*}_{i_{a}}$, where the last inequality holds by (\ref{optimal2}).
	Then, by  $z^{*}_{i_{a}}\leq x^{\rm f}_i<x^*_i=z^*_{i_{a}}+z^*_{i_{b}}$ and 
    $z^*_{i_{b}}\in \{0,1\}$ (Proposition~\ref{optimal}), we have 
    $z^*_{i_{b}}=1$. 
    By Lemma~\ref{prepare_optimal} (ii), 
    we have $z^{*}_{i_{a}}=\left\lfloor  B_i/v_i \right\rfloor$.
    Moreover, by $\left\lfloor  B_i/v_i \right\rfloor \leq x^{\rm f}_i<\left\lfloor  B_i/v_i \right\rfloor+1$ and $x^{\rm f}_i\in \mathbb Z_+$, it holds 
    $x^{\rm f}_i=\left\lfloor  B_i/v_i \right\rfloor$. Therefore, we have 
	$x^{\rm f}_i=z^{*}_{i_{a}}=\left\lfloor  B_i/v_i \right\rfloor$~and~$z^*_{i_{b}}=1$.

	Moreover, suppose to the contrary that $c^{\rm f}_i<v_{i_{b}}$. 
	By (\ref{x}), we have 
        $p^{\rm f}_i\leq \left\lceil p^{\rm f}_i/c^{\rm f}_i\right\rceil c^{\rm f}_i\leq c^{\rm f}_i x^{\rm f}_i= \left\lfloor  B_i/v_i \right\rfloor c^{\rm f}_i$.
	Then, by the definition of $v_{i_{b}}$ and $c^{\rm f}_i<v_{i_b}<v_i$, we have
	$B_i-p^{\rm f}_i=\left\lfloor B_i/v_i \right\rfloor v_i+v_{i_{b}}-p^{\rm f}_i \geq  \left\lfloor B_i/v_i \right\rfloor v_i+v_{i_{b}}-
	\left\lfloor  B_i/v_i \right\rfloor c^{\rm f}_i
	   > v_{i_{b}}>c^{\rm f}_i$,
	contradicting with $(B_i-p^{\rm f}_i)/c^{\rm f}_i\leq 1$. Thus, we have~$v_{i_{b}}\leq c^{\rm f}_i$.
	\end{proof}

	\begin{proof}[Proof of Lemma \ref{prepare}]
        (i) Already proved in Lemma \ref{prepare_optimal} (iii).

	(ii) If $x^{\rm f}_i\geq x^*_i$ or $c^{\rm f}_i=v_i$, 
        then by $v_{i_{b}}\leq v_i=v_{i_{a}}$, it holds 
	$v_i x^{\rm f}_i-(v_{i_{a}} z^{*}_{i_{a}}+v_{i_{b}} z^*_{i_{b}})\geq v_i (x^{\rm f}_i-x^*_i)\geq c^{\rm f}_i(x^{\rm f}_i-x^*_i)$.
	If $x^{\rm f}_i< x^*_i$ and $c^{\rm f}_i<v_i$, by Lemma \ref{algo<opt}, 
	it holds $x^{\rm f}_i=z^{*}_{i_{a}}=\left\lfloor  B_i/v_i \right\rfloor$,
	$z^*_{i_{b}}=1$, and $v_{i_{b}}\leq c^{\rm f}_i$.
	Thus,  
	it holds $v_i x^{\rm f}_i-(v_{i_{a}} z^{*}_{i_{a}}+v_{i_{b}} z^*_{i_{b}})=-v_{i_{b}}z^*_{i_{b}}\geq -c^{\rm f}_{i}z^*_{i_{b}}
	=c^{\rm f}_i(x^{\rm f}_i-x^*_i)$.

	(iii) Consider an arbitrary buyer $i$ with $x_i^{\rm f} < x_i^*$. Suppose that $c^{\rm f}_i<v_i(=v_{i_a})$.
	By Lemma \ref{algo<opt}, we have  
	$x^{\rm f}_i=z^{*}_{i_{a}}=\left\lfloor  B_i/v_i \right\rfloor$, $z^*_{i_{b}}=1$, and $v_{i_{b}}\leq c^{\rm f}_i$. Combining this with Lemma \ref{prepare_optimal} (iii) and $v_i=v_{i_a}$, we have~$\min(v_i x^*_i, B_i)-\min(v_i x^{\rm f}_i, B_i)
 =v_{i_a}z_{i_{a}}^*+v_{i_{b}}z_{i_{b}}^*-v_i x^{\rm f}_i=v_{i_{b}}z_{i_{b}}^*=
	c^{\rm f}_{i}z_{i_{a}}^*+v_{i_{b}}z_{i_{b}}^*-c^{\rm f}_{i}x^{\rm f}_{i}.$
	Suppose that $c^{\rm f}_i=v_i(=v_{i_a})$.
        If $B_i\geq v_i x^{\rm f}_i$, then by (i), it holds 
        $\min(v_i x^*_i, B_i)-\min(v_i x^{\rm f}_i, B_i)
        = v_{i_a}z_{i_a}^*+v_{i_b}z_{i_b}^*-v_i x^{\rm f}_{i}
	= c^{\rm f}_{i}z_{i_{a}}^*+v_{i_{b}}z_{i_{b}}^*-c^{\rm f}_{i}x^{\rm f}_{i}.$
        If $B_i<v_i x^{\rm f}_i(<v_i x^*_i)$, 
        then by (\ref{optimal2}), it holds $x^*_i\geq x^{\rm f}_i+1> B_i/v_i +1\geq z^{*}_{i_a}+z^{*}_{i_b}$, which means that this case never happens.
        Therefore, it holds 
	$\min(v_i x^*_i, B_i)-\min(v_i x^{\rm f}_i, B_i)=c^{\rm f}_{i}z_{i_{a}}^*+v_{i_{b}}z_{i_{b}}^*-c^{\rm f}_{i}x^{\rm f}_{i}$ 
	for each $i$ with $x^{\rm f}_i< x^*_i$.
	\end{proof}
	
\subsubsection{Proof of Lemma \ref{payment}}
We first show a lower bound on the number of goods remaining at any point in our mechanism. 
Consider an iteration of Algorithm 1, where $x$ and $d$ are the allocation of goods and the demand vector, respectively, and $c$ is the price.
Define 
\begin{equation}
\label{xandy}
 X:=\{i\in N\mid d_i>0\}\ \ {\rm and}\ \ 
 Y:=\{i\in X\mid x^{\rm f}_i< x^*_i\}
\end{equation}
 as in Sato \cite{S2023}.
In addition, we define a new set
\begin{equation}
\label{yc}
 Y_c:=\{i\in Y\mid c<v_{i_{b}}\}.
\end{equation}
 Obviously, it holds $Y_c\subseteq Y\subseteq X$.
 Moreover, define $z^*_a:=(z^*_{i_a})_{i\in N}$ and $z^*_b:= (z^*_{i_b})_{i\in N}$.
 Then, the following holds:
 
	\begin{lemma}
	\label{opt_algo_relation}
	It holds $f_{x,d}(Y)\geq z_a^*(Y)+z_b^*(Y_c)-x(Y)$.
	\end{lemma}
         If $Y_c$ is replaced by $Y$,
    this inequality is changed to $f_{x,d}(Y)\geq x^*(Y)-x(Y)$,   
	which is the same as Theorem 3.9 of Sato \cite{S2023}. 
However, due to the indivisibility, the number of remaining goods might be fewer than this bound.

	To handle with the difference, 
	we used the optimal allocation of virtual buyers $z^*_a$ and $z^*_b$ instead of $x^*$ 
	and defined a new set $Y_c$ to obtain this sharp lower bound. 
 
	\begin{proof}
	Let $Y'\subseteq Y$ be a minimizer of $f_{x,d}(Y)$, i.e., 
	$f_{x,d}(Y)=f(Y')-x(Y')+d(Y\setminus Y')$
	by Lemma~\ref{fxd_properties}~(ii). 
	We define $\lambda:=(\lambda_i)_{i\in N}$ by 
	\[
	\lambda_i=
	\begin{cases}
	1&\ \  {\rm if}\ \ c< v_{i_{b}}\\
	0&\ \  {\rm otherwise}
	\end{cases}
        \qquad (i\in N).
	\] 
        Then, by $z^*_{i_b}\in\{0,1\}$ for each $i\in N$ (from (\ref{optimal2})),
        we have $z^*_{b}(Y_c\setminus Y')\leq |Y_c\setminus Y'|=\lambda(Y\setminus Y')$.
        Using this, we have  
	\begin{align*}
	&f_{x,d}(Y)-z_a^*(Y)-z_b^*(Y_c)+x(Y)\\
	&\quad=f(Y')-x(Y')+d(Y\setminus Y')-z_a^*(Y)-z_b^*(Y_c)+x(Y)\\
	&\quad=f(Y')-z_a^*(Y')-z_b^*(Y'\cap Y_c)+x(Y\setminus Y')+d(Y\setminus Y')-z_a^*(Y\setminus Y')-z_b^*(Y_c\setminus Y')\\
	&\quad\geq x(Y\setminus Y')+d(Y\setminus Y')-z_a^*(Y\setminus Y')-z_b^*(Y_c\setminus Y')
	\geq\sum_{i\in Y\setminus Y'}(x_i+d_i-z^*_{i_a}-\lambda_i), 
	\end{align*}
	where the first inequality holds by polymatroid constraint 
	$f(Y')-z_a^*(Y')-z_b^*(Y'\cap Y_c)\geq f(Y')-x^*(Y')\geq 0$, 
	and the second inequality holds by 
	$z^*_{b}(Y_c\setminus Y')\leq \lambda(Y\setminus Y')$.
 
	In the following, we show $x_i+d_i-z^{*}_{i_{a}}-\lambda_i\geq 0$ for each $i\in Y\setminus Y'$.
        Since buyer $i\in Y\setminus Y'$ is active, it holds $c\leq v_i$.
        Since buyers clinch their allocated goods at a price less than or equal to the current price, it also holds $x_i\geq \left\lceil  p_i/c\right\rceil$.
        
	Case 1: Suppose that the demand $d_i$ has never been updated in line 9. 
	By Lemma \ref{d} (i) and the polymatroid constraint, 
	we have $x_i+d_i-z^{*}_{i_{a}}-\lambda_i= f(i)+1-z^{*}_{i_{a}}-\lambda_i\geq 0$.
 
	Case 2: Suppose that the demand $d_i$ has been updated in line 9 at least once.
        By Lemma \ref{d} (ii) and (iii), 
        it holds $d_i= (B_i-p_i)/c-1$ only when $c<v_i$ and $ (B_i-p_i)/c\in \mathbb Z_{+}$ 
        (see line 9 of Algorithm 1).
        If this is not the case, by Lemma \ref{d} (iii), it holds $d_i=\left\lfloor(B_i-p_i)/c\right\rfloor$ and $c\leq v_i$.
        Then, in both cases, we have  
        $d_i \geq \left\lfloor (B_i-p_i)/v_i\right\rfloor$ and
        $d_i \geq \left\lfloor (B_i-p_i)/c'\right\rfloor$ for any $c'>c$.
        
        Case 2-1: $v_{i_{b}}\leq c (\leq v_i)$.  
        In this case, it holds  $\lambda_i=0$.
        Moreover, it holds $d_i\geq \left\lfloor (B_i-p_i)/v_i\right\rfloor
        \geq \left\lfloor B_i/v_i \right\rfloor -\left\lceil p_i/v_i\right\rceil$.
        Combining the above with $x_i\geq \left\lceil  p_i/c\right\rceil\geq \left\lceil  p_i/v_i\right\rceil$ and $z^{*}_{i_{a}}\leq B_{i_a}/v_{i_a}= \left\lfloor B_i/v_i \right\rfloor$ (by (\ref{optimal2})), we have
	\[
	x_i+d_i-z^{*}_{i_{a}}-\lambda_i\geq \left\lceil  p_i/v_i\right\rceil+\left\lfloor  B_i/v_i \right\rfloor-\left\lceil  p_i/v_i\right\rceil- \left\lfloor  B_i/v_i \right\rfloor=0.
	\]

        Case 2-2: $c< v_{i_{b}} (< v_i)$.
         In this case, it holds $\lambda_i=1$.
    Moreover, by $v_{i_{b}}>c$ and $d_i \geq \left\lfloor (B_i-p_i)/c'\right\rfloor$ for any $c'>c$, it holds 
	\begin{align*}
        d_i&\geq\left\lfloor (B_i-p_i)/v_{i_{b}}\right\rfloor=
	\left\lfloor \frac{\left\lfloor  B_i/v_i \right\rfloor v_i+v_{i_{b}}-p_i}{v_{i_{b}}}\right\rfloor
        \geq\left\lfloor \frac{\left\lfloor  B_i/v_i \right\rfloor v_{i_b}+v_{i_{b}}-p_i}{v_{i_{b}}}\right\rfloor\\
	&=\left\lfloor \left\lfloor  B_i/v_i \right\rfloor +1-\frac{p_i}{v_{i_{b}}}\right\rfloor=
 \left\lfloor  B_i/v_i \right\rfloor +1-\left\lceil p_i/v_{i_{b}}\right\rceil,
       \end{align*}
       where the first equality holds by the definition of $v_{i_b}$ and the second inequality holds by $v_{i_b}<v_i$.
        Combining the above with $x_i\geq \left\lceil  p_i/c\right\rceil\geq \left\lceil  p_i/v_{i_b}\right\rceil$ and $z^{*}_{i_{a}}\leq B_{i_a}/v_{i_a}= \left\lfloor B_i/v_i \right\rfloor$, we have 
	\[
	x_i+d_i-z^{*}_{i_{a}}-\lambda_i\geq
	\left\lceil p_i/v_{i_{b}}\right\rceil+ \left\lfloor  B_i/v_i \right\rfloor+1-\left\lceil p_i/v_{i_{b}}\right\rceil- \left\lfloor  B_i/v_i \right\rfloor-1=0.
	\]
	\end{proof}

	Since the remaining goods are sold at the price more than the current one, we provide a lower bound of future payments in our mechanism for the set of buyer $i$ 
	with $x^{\rm f}_i\geq x^*_i$.
	As in Lemma~\ref{opt_algo_relation}, 
	we use the optimal allocation of virtual buyers instead of 
	that of original buyers.
	In the proof, we use {\it backward} mathematical induction as in the proof of Theorem 5.9 in Sato~\cite{S2023}.

	\begin{proof}[Proof of Lemma \ref{payment}]
	Recall that $Y_c\subseteq Y\subseteq X$ defined in (\ref{xandy}) and (\ref{yc}). We show that, throughout Algorithm 1, it holds
	\begin{equation}
	\label{payment2}
	\sum_{i\in X\setminus Y}(p^{\rm f}_i-p_i)\geq\sum_{i\in Y}c^{\rm f}_{i}(z_{i_{a}}^*-x^{\rm         f}_{i})+\sum_{i\in Y_c}v_{i_{b}}z_{i_{b}}^*+c(f_{x,d}(X)-z_a^*(Y)-z_b^*(Y_c)+x(Y)).
	\end{equation} 
    If (\ref{payment2}) holds, then we consider the beginning of the auction. By $X=N$ and $Y=Y_c=\{i\in N: x^{\rm f}_i<x^*_i\}$, we have 
	$\sum_{i\in N: x^{\rm f}_i\geq x^*_i}p^{\rm f}_i\geq \sum_{i\in N: x^{\rm f}_i<x^*_i}
	(c^{\rm f}_{i}z_{i_{a}}^*+v_{i_{b}}z_{i_{b}}^*-c^{\rm f}_{i}x^{\rm f}_{i})$, as required.
	
    For proving (\ref{payment2}), we use 
	\begin{align}
    \label{f_xd_X}
	f_{x,d}(X)+x(Y)=f(N)-x(N\setminus Y),
	\end{align}
	which is obtained by $f_{x,d}(X)=f_{x,d}(N)$ and $f_{x,d}(N)=f(N)-x(N)$. 
    Note that $f_{x,d}(X)=f_{x,d}(N)$ holds by $f_{x,d}(X)\leq f_{x,d}(N)$ (the monotonicity of $f_{x,d}$)
	and $f_{x,d}(N)\leq f_{x,d}(X)+d(N\setminus X)=f_{x,d}(X)$ by (\ref{naive}) and the definition of $X$.
    Also, $f_{x,d}(N)=f(N)-x(N)$ holds throughout Algorithm 1; see the proof of Theorem \ref{all_goods}.
    Then, substituting this for (\ref{payment2}), our goal is also expressed by  
    \begin{equation}
	\label{payment3}
	\sum_{i\in X\setminus Y}(p^{\rm f}_i-p_i)\geq\sum_{i\in Y}c^{\rm f}_{i}(z_{i_{a}}^*-x^{\rm         f}_{i})+\sum_{i\in Y_c}v_{i_{b}}z_{i_{b}}^*+c(f_{x,d}(X)-z_a^*(Y)-z_b^*(Y_c)+x(Y)).
	\end{equation} 

    We show that if (\ref{payment2}) (or (\ref{payment3})) holds at the end of an iteration, it holds at the beginning of the iteration.
	At the end of Algorithm 1, (\ref{payment2}) holds in equality since the both sides are equal to $0$ by $X=Y=Y_c=\emptyset$.
        Now we perform the following case-by-case analysis.
	    Note that we only focus on active buyers $i\in X$ since (\ref{payment2}) is only influenced by the changes of active buyers.

	(i) Execution of Algorithm 2: For buyer $i$ who belongs to $X\setminus Y$ just before clinching $\delta_i\geq 0$ amount of goods, 
	the left-hand side of (\ref{payment2}) is decreased by $c\delta_i$.
	The first and the second terms on the right-hand side are unchanged by $i\notin Y$ (before and after the clinching), and
	the third term on the right side is decreased by $c\delta_i$ 
	by $f_{x,d}(X)+x(Y)=f(N)-x(N\setminus Y)$ (=(\ref{f_xd_X})).
	Thus, both sides of (\ref{payment2}) are decreased by $c\delta_i$.
	
        Suppose that $i$ belongs to $Y$ just before the clinching.
        Since $i$ keeps active at that time, it holds $c\leq v_i$.
        Then, after the clinching step, it holds $i\notin X\setminus Y$ whether $i$ is active or not.
        Therefore, the left-hand side is unchanged by the clinching.
        If $i$ is still active, i.e., $i\in Y$
	the first and the second terms on the right-hand side are unchanged.
	Moreover, the third term on the right side is also unchanged by (\ref{f_xd_X}).
	Thus, both sides of (\ref{payment2}) are unchanged.
        If the buyer $i$ is dropping by clinching, 
        it must hold $i\in Y\setminus Y_c$. This can be shown by the following:
	If $c=v_i$, then $v_{i_b}<v_i=c$.  
	If $c<v_i$, by $x^{\rm f}_i<x^*_i$ (from $i\in Y$) and $c^{\rm f}_i=c<v_i$, we can apply Lemma \ref{algo<opt}, 
        and thus we have $v_{i_b}\leq c^{\rm f}_i=c$.
        After the clinching, since $i$ is dropping, $i$ does not belong to any of $X, Y, Y_c$. 
        Then, on the right-hand side, the first term is decreased by $c^{\rm f}_i (z^*_{i_a}-x^{\rm f}_i)=c(z^*_{i_a}-x^{\rm f}_i)$, 
	and the second term is unchanged by $i\notin Y_c$ before and after the clinching, 
	and the third term is increased by $c(z^*_{i_a}-x^{\rm f}_i)$ from (\ref{f_xd_X}).
	Again, both sides of (\ref{payment2}) are unchanged. 
	Therefore, if (\ref{payment2}) holds after the execution of Algorithm 2, 
	it holds before that.
	
	(ii) The demand update: Suppose that the demand $d_i$ of buyer $i$ is updated in line 5 or 9.
	If the demand $d_i$ is still positive after the update, 
	both sides of (\ref{payment2}) are unchanged 
	since the right-hand side of (\ref{f_xd_X}) is unchanged.
	In the following, we consider the case where the demand $d_i$ becomes zero.
        Then, after the update, $i$ does not belong to any of $X, Y, Y_c$. 
	The left-hand side of (\ref{payment2}) is unchanged 
        since it holds $p_i=p^{\rm f}_i$ even before the update.
	If $i\in X\setminus Y$ just before the update, then the right-hand side is unchanged by (\ref{f_xd_X}).
	If $i\in Y_c$, then it holds $c^{\rm f}_i=c<v_{i_b}<v_i$, which never happens by $x^{\rm f}_i<x^*_i$ (from $i\in Y$) and Lemma~\ref{algo<opt}.
        If $i\in Y\setminus Y_c$, 
	then the first term on the right-hand side is decreased by $c^{\rm f}_i(z^*_{i_a}-x^{\rm f}_i)=c(z^*_{i_a}-x^{\rm f}_i)$, 
	the second term is unchanged, 
	and the third term is increased by $c(z^*_{i_a}-x_i)=c(z^*_{i_a}-x^{\rm f}_i)$ (from (\ref{f_xd_X})).
	Thus, the right-hand side is also unchanged.
	Therefore, if (\ref{payment2}) holds after the demand update, it holds before that.
	
	(iii) The price update: It suffices to consider the change of $Y_{c}=\{i\in Y\mid c<v_{i_b}\}$ in the second and third terms on the right-hand side of (\ref{payment2}) 
	since $(x,p)$ and $d$ are unchanged by the price update.
	Let $\tilde{c}$ be the price before the update. We consider the change in the right-hand side of (\ref{payment2}) when the price increases from $\tilde{c}$ to $c$. 
	Then, we have
	\begin{align*}
	\sum_{i\in Y_c}&v_{i_{b}}z_{i_{b}}^*+c(f_{x,d}(X)-z_a^*(Y)-z_b^*(Y_c)+x(Y))\\
	&=\bigl(\sum_{i\in Y_c}v_{i_{b}}z_{i_{b}}^*+c z_b^*(Y_{\tilde{c}}\setminus Y_c)\bigr)+c(f_{x,d}(X)-z_a^*(Y)-z_b^*(Y_{\tilde{c}})+x(Y)) \\
	&\geq\sum_{i\in Y_{\tilde{c}}}v_{i_{b}}z_{i_{b}}^*+c(f_{x,d}(X)-z_a^*(Y)-z_b^*(Y_{\tilde{c}})+x(Y))\\
 	&\geq\sum_{i\in Y_{\tilde{c}}}v_{i_{b}}z_{i_{b}}^*+\tilde{c}(f_{x,d}(X)-z_a^*(Y)-z_b^*(Y_{\tilde{c}})+x(Y)),
	\end{align*}
	where the first inequality holds by $(\tilde{c}<) v_{i_b}\leq c$ for each $i\in Y_{\tilde{c}}\setminus Y_{c}$ and $z_b^*(Y_{\tilde{c}}\setminus Y_{c})\geq 0$ (by $Y_{\tilde{c}}\supseteq Y_{c}$) and the last inequality holds by $\tilde{c}<c$ and $f_{x,d}(Y)-z_a^*(Y)-z_b^*(Y_{\tilde{c}})+x(Y)\geq 0$ (Lemma~\ref{opt_algo_relation}).
	Thus, the left-hand side of (\ref{payment2}) is unchanged, and the 
	right-hand side is increased after the price update. By the backward induction, if (\ref{payment2}) holds for new price $c$, it also holds for the old price $\tilde{c}$.
	
    Therefore, we conclude that (\ref{payment2}) holds throughout Algorithm 1.
	\end{proof}

\section{Concluding Remarks}
	In this study, we propose the polyhedral clinching auction for indivisible goods, 
	which satisfies IC and IR, and works with polymatroidal environments. 
	Moreover, we provided the tight sets lemma and three types of efficiency guarantees (PO, LW, SW). 
	Since many of our results have not been shown even in special cases of our setting, we believe that our results significantly advance the efficiency guarantee of clinching auctions for indivisible goods. Our results are also helpful in extending clinching auctions to two-sided markets. These results will be included in the revised version of Sato \cite{S2023}.

	We note some possible future directions of our study.
	Firstly, there still remains a gap between the approximation ratio and the lower bound for LW guarantees, 
	which can be seen by Proposition~\ref{example}.
	In the divisible setting, Lu and Xiao \cite{LX2015} proposed a mechanism 
	that satisfies IC and IR and achieves a better approximation ratio than clinching auctions. 
	Seeking for such mechanisms in the indivisible settings, 
	especially for polymatroidal environments, is an interesting future work.
	
    Secondly, it may be of particular interest to find another type of SW guarantee specific to the indivisible settings.
    Devanur et al.~\cite{DHH2013} showed that in their divisible setting, clinching auctions yield an envy-free allocation and achieve a 2-approximation of SW to the maximum SW among all envy-free allocations.
    In our setting, envy-freeness can be defined in the same way as theirs:  
	\begin{definition}[e.g., Devanur et al. \cite{DHH2013}]
	An allocation $\mathcal A$ is envy-free if it holds 
	$u_i(\mathcal A)\geq u_i(\mathcal A_{ij})$ for each $i,j\in N$, 
	where $\mathcal A_{ij}$ is the allocation obtained from $\mathcal A$ by swapping the allocation of $i$ and~$j$.
	\end{definition}
    Extending their results to our indivisible setting seems to be a potential candidate for another SW guarantee.
	However, this type of guarantees seems unrealistic due to the indivisibility even if we use an approximate notion of envy-freeness. 
    This can be seen in the following example:
	
\begin{example}
	\label{envy_free}
	The seller owns one unit of an indivisible good. 
	Buyer 1 has valuation $k\ (k\gg 2)$ and buyer 2 has valuation $2$, 
	and they have a common budget of $1$. 
	In Algorithm 1, at $c=1$, the demand $d_1$ decreases to $0$ in line 9, and buyer 2 clinches one unit of goods at price $1$. 
	Then, we have $(x^{\rm f}, p^{\rm f})=((0,1), (0,1))$, for which 
	buyer 1 feels envy about the allocation of buyer 2.
	
	If we swap the valuations of buyer 1 and buyer 2, the allocation is unchanged.
	This implies that if we use some approximate notion of envy-freeness, 
	we must use the one that admits both $(x^{\rm f}, p^{\rm f})=((0,1), (0,1))$ and $(x^{\rm f}, p^{\rm f})=((1,0), (1,0))$.
	Then, the SW for $(x^{\rm f}, p^{\rm f})=((0,1), (0,1))$ and $(x^{\rm f}, p^{\rm f})=((1,0), (1,0))$ is $2$ and $k$, respectively.
	Therefore, the approximation ratio of SW is at most $2/k$, which reaches zero as we take $k\to\infty$.
\end{example}
Therefore, it would be interesting to examine what benchmarks other than LW and envy-free benchmarks might be useful for SW guarantees in budget constrained auctions.

\section*{Acknowledgement}
	A preliminary version of this paper has appeared in Proceedings of the 19th Conference on Web and Internet Economics, Shanghai, China, December 2023.
	We thank anonymous reviewers for helpful feedback and suggestions.
	This work was supported by Grant-in-Aid for JSPS Research Fellow Grant Number JP22KJ1137,  
	Grant-in-Aid for Challenging Research (Exploratory) Grant Number JP21K19759, and JST ERATO Grant Number JPMJER2301.

\bibliography{Indivisible}
\appendix

\section{Relation to No Trading Path Property}
Here we show that our tight sets lemma implies the
{\it no trading path property}. 
Fiat et al. \cite{FLSS2011}  and Colini-Baldeschi et al. \cite{BHLS2015} 
showed that in some special cases of our setting, the non-existence of a trading path and selling all goods are equivalent to PO, and that
their clinching auction satisfies this property.
Thus, we used a different basis of efficiency from theirs, and it would be interesting to examine the relationship between these two properties.
	
    First, we introduce a {\it trading pair}, an extension of their trading path to polymatroidal environments. 
    For preparation, following the notations in Fujishige \cite{F2005}, we define the saturation and dependence functions.
	A saturation function ${\rm sat}:P \to 2^N$ is defined by 
	${\rm sat}(x):=\{i\mid i\in N, \forall \alpha>0, x+\alpha \chi_i\notin P\}$,
	where $\chi_{A}$ represents an indicator vector for $A\subseteq N$.
	Moreover, for $x\in P$ and $i \in {\rm sat}(x)$, 
	a dependence function ${\rm dep} : P \times N \to 2^N$ is defined by
	${\rm dep}(x,i):=\{i'\mid i'\in N, \exists \alpha>0, x+\alpha (\chi_i-\chi_{i'})\in P\}$.

\begin{definition}
\label{trading_pair}
{\rm A pair of buyers $(i, j)$ is a} trading pair {\rm with respect to the allocation
$(x, p)$ if the following hold: (i)\, $j\in {\rm dep}(x, i)$, (ii)\, $v_i$ is strictly greater than $v_j$, and 
(iii) the remaining budget $B_i-p_i$ is not less than $v_j$.}
\end{definition}

    Intuitively, if there exists a trading pair $(i,j)$, 
    then by selling $j$'s goods to $i$ at the price $v_j$, 
    buyer~$i$ can improve her utility without changing the utility of other buyers. 
    This means that the allocation is not Pareto optimal.

    Suppose that the polymatroid $P$ is induced from a bipartite graph 
    whose two vertex sets consist of the buyers and the goods, respectively. 
    Let $(i,i')$ be a trading pair. 
    Then, there exists a path from $i$ to $i'$ on this graph,
    which is exactly the trading path in the sense of Fiat et al. \cite{FLSS2011}  and Colini-Baldeschi et al. \cite{BHLS2015}.
   In this way, our trading pair is a generalization of their trading~path.

Using our tight sets lemma, we show that our mechanism satisfies the no trading pair property. 
Theorem \ref{no_trading} implies that our tight sets lemma is a stronger basis for efficiency.

\begin{theorem}
\label{no_trading}
There is no trading pair with respect to $(x^{\rm f}, p^{\rm f})$.
\end{theorem}
\begin{proof}
Let $\{X_k\}$ be the tight sets in Theorem \ref{tightsets} (i).
Suppose that there exists a trading pair $(i, j)$, and $i\in X_{k'}\setminus X_{k'-1}$ for some $k'\in [t]$.
By Theorem \ref{tightsets} (i), it holds that $x^{\rm f}(X_{k'})=f(X_{k'})$.
Thus, by $i\in X_{k'}\setminus X_{k'-1}$ and $j\in {\rm dep}(x, i)$, we have $j\in X_{k'}$.
Moreover, by Theorem \ref{tightsets} (ii), we have $v_j\geq c^{\rm f}_j\geq c^{\rm f}_{i_{k'}}=c^{\rm f}_i$.

Suppose that $v_{i}=c^{\rm f}_{i}$. Then, it holds $v_j\geq c^{\rm f}_j\geq c^{\rm f}_i=v_i$, 
contradicting Definition \ref{trading_pair} (ii). 
Suppose that $v_i>c^{\rm f}_i$.
By Theorem \ref{tightsets} (iii) and (iv) [(b) $\Rightarrow$ (c)], it holds $B_i-p^{\rm f}_i\leq c^{\rm f}_i$.
Thus, if $v_j>c^{\rm f}_j$, it holds that $B_i-p^{\rm f}_i\leq c^{\rm f}_i\leq c^{\rm f}_j<v_j$, which contradicts Definition \ref{trading_pair} (iii). 

The remaining case is $v_i>c^{\rm f}_i$ and $v_j=c^{\rm f}_j$.
Suppose that $c^{\rm f}_j>c^{\rm f}_i$.
Then, we have $B_i-p^{\rm f}_i\leq c^{\rm f}_i< c^{\rm f}_j=v_j$, which contradicts Definition~\ref{trading_pair}~(iii).
Suppose that $c^{\rm f}_i=c^{\rm f}_j$. Then, by $i\in X_{k'}\setminus X_{k'-1}$ and $j\in X_{k'}$, 
buyer $i$ was dropped before or at the same moment in the iteration where $j$ dropped. 
Since buyer $j$ was dropped in line 5 or line 6 of Algorithm 1 by $v_j=c^{\rm f}_j$, 
buyer $i$ was dropped in line 6 of Algorithm 1 by $v_i>c^{\rm f}_i$. 
By Theorem~\ref{tightsets} (iv) $[\neg (b) \Rightarrow \neg (a)]$, 
this implies $(B_i-p^{\rm f}_i)/c^{\rm f}_i<1$.
Then, again, we have $B_i-p^{\rm f}_i<c^{\rm f}_i=c^{\rm f}_j=v_j$, 
which contradicts Definition~\ref{trading_pair} (iii).
\end{proof}

\section{Omitted Proofs}
\subsection{Proof of Lemma \ref{clinch_amount}}
We first prove Lemma \ref{clinch_amount}.
Although we use almost the same idea as Goel et al. \cite{GMP2015}, their proof omits some technical arguments.
Therefore, we now provide the full proof for completeness. In the proof, we use the following lemmas:
\begin{lemma}
\label{naive_property}
In (\ref{naive}), for each $S\subseteq N$, there exist $S^*\subseteq S$ and $S^{\dag}\supseteq S^*$ with $S^{\dag}\cap (S\setminus S^*)=\emptyset$ such that $f_{x,d}(S)=f(S^{\dag})-x(S^{\dag})+d(S\setminus S^*)$.
\end{lemma}
\begin{proof}
By (\ref{naive}), there exists $S^*\subseteq S$ and $S^{\dag}\supseteq S^*$ such that $f_{x,d}(S)=f(S^{\dag})-x(S^{\dag})+d(S\setminus S^*)$.
    If $(S^{\dag}\setminus S^*)\cap S\neq\emptyset$, define $\tilde{S}:=(S^{\dag}\setminus S^*)\cap S$.
    Then, it holds $S^*\cup \tilde{S}\subseteq S$ and 
    $S^{\dag}\supseteq S^*\cup \tilde{S}$.
    Moreover, it holds $f(S^{\dag})-x(S^{\dag})+d(S\setminus (S^*\cup \tilde{S}))\leq f(S^{\dag})-x(S^{\dag})+d(S\setminus S^*)=f_{x,d}(S)$.
    By (\ref{naive}), this inequality holds in equality.
    Then, we can replace $S^*$ with $S^*\cup \tilde{S}$. 
    Therefore, in (\ref{naive}), we can take $S^*\subseteq S$ 
and $S^{\dag}\supseteq S^*$ such that $S^{\dag}\cap (S\setminus S^*)=\emptyset$.
\end{proof}

\begin{lemma}
\label{fxdw}
The monotone submodular function $f^{i, w_i}_{x,d}: 2^{N\setminus i}\to \mathbb Z_{+}$ that defines $P^i_{x,d}(w_i)$ is represented by 
$f^{i, w_i}_{x,d}(S):=\min\{f_{x,d}(S\cup i)-w_i, f_{x,d}(S)\}\ \ (S\subseteq N\setminus i)$.
\end{lemma}
\begin{proof}
By the definition of $P^i_{x,d}(w_i)$ (where $w_i\leq d_i$), it holds 
$f^{i, w_i}_{x,d}(S)=f_{\hat{x},d}(S)$\ \ $(S\subseteq N\setminus i)$, 
where $\hat{x}$ is obtained from $x$ by replacing $x_i$ with $x_i+w_i$.
By (\ref{naive}) and $x_i\leq \hat{x}_i\ (i\in N)$, we have 
\begin{equation}
\label{fhatxd}
f_{\hat{x},d}(S)=\min_{S'\subseteq S}\{\min_{S''\supseteq S'}
\{f(S'')-\hat{x}(S'')\}+d(S\setminus (S')\}\leq\min_{S'\subseteq S}\{\min_{S''\supseteq S'}\{f(S'')-x(S'')\}+d(S\setminus S'\}=f_{x,d}(S).
\end{equation}
By (\ref{naive}), there exist $S^*\subseteq S$ and $S^{\dag}\supseteq S^*$  such that $f_{\hat{x},d}(S)=f(S^{\dag})-\hat{x}(S^{\dag})+d(S\setminus S^*)$.
Now we perform the following case-by-case analysis:

Case 1: $i\in S^{\dag}$. Then, it holds 
$f_{\hat{x},d}(S)=f(S^{\dag})-\hat{x}(S^{\dag})+d(S\setminus S^*)=f(S^{\dag})-x(S^{\dag})+d(S\setminus S^*)-w_i$. Moreover, by (\ref{naive}) and $i\in S^{\dag}$, we have  
$f(S^{\dag})-x(S^{\dag})+d(S\setminus S^*)=f(S^{\dag})-\hat{x}(S^{\dag})+d(S\setminus S^*)+w_i=\min_{S'\subseteq S}\{\min_{S''\supseteq S'}\{f(S''\cup i)-\hat{x}(S''\cup i)\}+d(S\setminus S')\}+w_i=\min_{S'\subseteq S}\{\min_{S''\supseteq S'}\{f(S''\cup i)-x(S''\cup i)\}+d(S\setminus S')\}$.
Therefore, by (\ref{naive}), we have 
\begin{align*}
f_{x,d}(S\cup i)&=\min\Bigl(f_{x,d}(S)+d_i, \min_{S'\subseteq S}\{\min_{S''\supseteq S'}
\{f(S''\cup i)-x(S''\cup i)\}+d(S\setminus S'\}\Bigr)\\
&=\min\Bigl(f_{x,d}(S)+d_i, f(S^{\dag})-x(S^{\dag})+d(S\setminus S^*)\Bigr)\\
&=\min\Bigl(f_{x,d}(S)+d_i, f_{\hat{x},d}(S)+w_i\Bigr)=f_{\hat{x},d}(S)+w_i,
\end{align*} 
where the last equality holds by $(\ref{fhatxd})$ and $w_i\leq d_i$.
Therefore, we have $f_{\hat{x},d}(S)=f_{x,d}(S\cup i)-w_i$.

Case 2: $i\notin S^{\dag}$. Then, it holds $f(S^{\dag})-x(S^{\dag})+d(S\setminus S^*)=f(S^{\dag})-\hat{x}(S^{\dag})+d(S\setminus S^*)= f_{\hat{x},d}(S)\leq f_{x,d}(S)$, where the last inequality holds by (\ref{fhatxd}).
By (\ref{naive}), it also holds $f(S^{\dag})-x(S^{\dag})+d(S\setminus S')\geq f_{x,d}(S)$. Therefore, we have 
$f_{\hat{x},d}(S)=f_{x,d}(S)$.

By the above, we have 
$f^{i, w_i}_{x,d}(S)=\min\{f_{x,d}(S\cup i)-w_i, f_{x,d}(S)\}$ 
for each $S\subseteq N\setminus i$, as required.
\end{proof}

\begin{proof}[Proof of Lemma \ref{clinch_amount}]
We consider the clinching amount of buyer $i$.
Then, the clinching condition is described by $P^i_{x',d'}(w_i)=P^i_{x',d'}(0)$, which is equivalent to 
$f^{i, w_i}_{x',d'}(S)=f^{i, 0}_{x',d'}(S)$.
By Lemma~\ref{fxdw} and the monotonicity of $f_{x',d'}$, we have $f^{i, 0}_{x',d'}(S)=\min\{f_{x',d'}(S\cup i), f_{x',d'}(S)\}=f_{x',d'}(S)$.
Then, by Lemma \ref{fxdw}, the clinching condition means $f_{x',d'}(S)\leq f_{x',d'}(S\cup i)-w_i$ for each $S\subseteq N\setminus i$.
Therefore, by the definition of $\delta_i$ and the submodularity of $f_{x',d'}$, we have
$\delta_i=\min_{S\subseteq N\setminus i} \{f_{x',d'}(S\cup i)-f_{x',d'}(S)\}=f_{x',d'}(N)-f_{x',d'}(N\setminus i)$. 

Subsequently, we show that the clinching amount is independent of the order of buyers. By Lemma \ref{naive_property}, for each $S\subseteq N$, 
we can take $S'\subseteq S$ and $S''\supseteq S'$ with $S''\cap (S\setminus S')=\emptyset$ such that 
$f_{x',d'}(S)=f(S'')-x'(S'')+d'(S\setminus S')$.
If $S=N$, then by $S''\supseteq S'$ and $S''\cap (N\setminus S')=\emptyset$, it must hold $S''=S'$.
Therefore, we have 
\begin{equation}
\label{N}
f_{x',d'}(N)=\min_{S'\subseteq N}\{f(S')-x'(S')+d'(N\setminus S')\}.    
\end{equation}

Similarly, by $S''\supseteq S'$ and $S''\cap (N\setminus (S'\cup i))=\emptyset$, if $S=N\setminus i$, it must hold $S''=S'$ or $S''=S'\cup i$.
Therefore, we also have 
\begin{align}
\label{Nsetminusi}
f_{x',d'}(N\setminus i)&=\min_{S'\subseteq N\setminus i}
\{\min\bigl(f(S'\cup i)-x'(S'\cup i), f(S')-x'(S')\bigr)+d'((N\setminus i)\setminus S')\}\nonumber \\ 
&\geq \min\bigl(f_{x',d'}(N), \min_{S'\subseteq N\setminus i} 
\{f(S')-x'(S')+d'((N\setminus i)\setminus S')\}\bigr),
\end{align}
where the inequality holds by (\ref{naive}) for $f_{x',d'}(N)$.
By the monotonicity of $f_{x',d'}$, it holds $f_{x',d'}(N\setminus i)\leq f_{x',d'}(N)$. By (\ref{naive}) for $f_{x',d'}(N\setminus i)$, 
we also have 
$f_{x',d'}(N\setminus i)\leq\min_{S'\subseteq N\setminus i} 
\{f(S')-x'(S')+d'((N\setminus i)\setminus S')\}$.
Therefore, the last inequality holds in equality. 
Combining (\ref{N}) and (\ref{Nsetminusi}), we have
\begin{align*}
&f_{x',d'}(N)-f_{x',d'}(N\setminus i)\\
&=\max(0, \min_{S'\subseteq N}\{f(S')-x'(S')+d'(N\setminus S')\}-\min_{S''\subseteq N\setminus i}\{f(S'')-x'(S'')+d'((N\setminus i)\setminus S'')\}).
\end{align*}
Now consider two buyers $j$ and $k$ with $j< k$.
if buyer $j$ clinches $\delta_j$ amount of goods before $k$'s clinching, 
$x'_j$ is updated to $x'_j=x'_j+\delta_j$ and $d'_j$ is updated to $d'_j=d'_j-\delta_j$.
Then, both $f(S')-x'(S')+d'(N\setminus S')$ and 
$f(S'')-x'(S'')+d'((N\setminus k)\setminus S'')$ decrease by $\delta_j$ for each $S'\subseteq N$ and $S''\subseteq N\setminus k$.
This implies that $f_{x',d'}(N)-f_{x',d'}(N\setminus k)$ is unchanged by the clinching of buyers numbered before $k$.
Therefore, we have $\delta_i=f_{x',d'}(N)-f_{x',d'}(N\setminus i)=f_{x,d}(N)-f_{x,d}(N\setminus i).$
\end{proof}

\subsection{Proof of Lemma \ref{fxd_properties} (ii)}
	Here we show Lemma \ref{fxd_properties} (ii). 
	Although the idea is the same as in Sato \cite{S2023}, due to the difference in mechanisms, we give the proof for completeness.
	We begin by the following lemma:
	\begin{lemma}
	\label{prepare_inv}
	In Algorithm 2, it holds
	$\displaystyle \delta(T\setminus S) \leq f_{x,d}(T)-f_{x,d}(S)$ 
	for each $S,T\subseteq N$ such that $T\supseteq S$, 
	where $x$ and $d$ are the allocation of goods and the demand vector, respectively, just before Algorithm~2.
	\end{lemma}
	\begin{proof}
	Suppose that $T\setminus S=\{\ell_1,\ell_2,\ldots, \ell_q\}$ for some positive integer $q$.
	We consider the transaction $\delta_i$ of buyer $i\in T\setminus S$.
	By Lemma \ref{clinch_amount} and the submodularity of $f_{x,d}$, 
	we have $\delta_i = f_{x,d}(N)-f_{x,d}(N\setminus i)\leq  f_{x,d}(U)-f_{x,d}(U\setminus i)$ 
	for each $U\ni i$, and thus 
	\begin{align*}
	\delta(T\setminus S) &=\sum_{i\in T\setminus S}(f_{x,d}(N)-f_{x,d}(N\setminus i)) \\
	&\leq f_{x,d}(T)-f_{x,d}(T\setminus \ell_1)+f_{x,d}(T\setminus \ell_1)-
	\cdots +f_{x,d}(S\cup \ell_q)-f_{x,d}(S)=f_{x,d}(T)-f_{x,d}(S).
	\end{align*}
	\end{proof}
	
	\begin{proof}[Proof of Lemma \ref{fxd_properties} (ii)]
	At the beginning of the auction, the claim holds by Lemma \ref{fxd_properties} (i).
	By the following case-by-case analysis, we prove that 
	if 
    \begin{equation}
    \label{hypothesis}
    f_{x,d}(S)=\min_{S'\subseteq S}\{f(S')-x(S')+d(S\setminus S')\}
    \end{equation}
    holds for each $S\subseteq N$ 
	before each step of our mechanism, then it also holds after the step.
	
	(i) The price update: Since $x$ and $d$ are unchanged by the price update, 
	the equation trivially holds even after the update.
	
	 (ii) The demand update: Suppose that the demand of buyer $i$ changes from $d_i$ to $d'_i$.
	Let $d'$ be the demand vector after the demand update of $i$, 
        i.e., the vector obtained from $d$ by replacing $d_i$ with $d'_i$.
	Then, in (\ref{naive}) for $f_{x,d'}(S)$, if $i\in S''$, then it holds $f_{x,d'}(S)=f_{x,d}(S)$.
	If $i\notin S''$, then it holds $f_{x,d'}(S)=f_{x,d}(S\setminus i)+d'_i$.
	Therefore, after the demand update, it holds 
	\[
	f_{x,d'}(S)=\min\{f_{x,d}(S),f_{x,d}(S\setminus i)+d'_i\}=
	\min_{S'\subseteq S}\{f(S')-x(S')+d'(S\setminus S')\} 
	\]
	by (\ref{hypothesis}) and 
	$f_{x,d}(S\setminus i)+d'_i<f_{x,d}(S\setminus i)+d_i$.

	(iii) Clinching Step: Let $x$ and $d$ (resp. $\tilde{x}$ and $\tilde{d}$) 
	be the allocation of goods and the demand vector, respectively, 
	just after (resp. before) the execution of Algorithm 2. 
	By (\ref{hypothesis}), 
	it holds $f_{\tilde{x},\tilde{d}}(S)=\min_{S'\subseteq S}\{f(S')-\tilde{x}(S')+\tilde{d}(S\setminus S')\}$.
	if buyer $i\in S$ clinches $\delta_i$ unit of goods in Algorithm 2,
	this increases $\tilde{x}_i$ by $\delta_i$ and decreases $\tilde{d}_i$ by $\delta_i$.
	Then, $\min_{S'\subseteq S}\{f(S')-\tilde{x}(S')+\tilde{d}(S\setminus S')\}$ 
	was decreased by $\delta(S)$ whether the buyers 
	belong to $S'$ or $S\setminus S'$. 
	This means that
	\[
	\min_{S'\subseteq S}\{f(S')-x(S')+d(S\setminus S')\}=\min_{S'\subseteq S}
	\{f(S')-\tilde{x}(S')+\tilde{d}(S\setminus S')\}-\delta(S)=f_{\tilde{x},\tilde{d}}(S)-\delta(S).
	\] 

  Suppose to the contrary that the equation in Lemma \ref{fxd_properties} (ii) does not hold just after Algorithm~2 has finished.
	By Lemma \ref{naive_property}, there exist $S^*\subseteq S$ and $S^{\dag}\supseteq S^*$ with $S^{\dag}\cap (S\setminus S^*)=\emptyset$ such that $f_{x,d}(S)=f(S^{\dag})-x(S^{\dag})+d(S\setminus S^*)$.
 Then, it holds  
	\begin{eqnarray}
	\label{X_aft}
	f_{x,d}(S)=f(S^\dag)-x(S^\dag)+d(S\setminus S^*)
	<\min_{S'\subseteq S}\{f(S')-x(S')+d(S\setminus S')\}=f_{\tilde{x},\tilde{d}}(S)-\delta(S).
	\end{eqnarray}
	By $S^*\subseteq S$, it holds $f(S^*)-x(S^*)+d(S\setminus S^*)\geq \min_{S'\subseteq S}\{f(S')-x(S')+d(S\setminus S')\}$. Then, the inequality in (\ref{X_aft}) implies $S^{\dag}\supset S^*$.
 By $S^{\dag}\supset S^*$ and $S^{\dag}\cap (S\setminus S^*)=\emptyset$, it holds $S^{\dag}\setminus S=S^{\dag}\setminus S^*\neq \emptyset$.
        Then, we have $S^\dag\cup S=(S^\dag\setminus S)\cup S\supset S$.
    Moreover, it holds
	\begin{align}
	\label{X_bef}
	f_{x,d}(S)&=f(S^\dag)-x(S^\dag)+d(S\setminus S^*) 
	\geq \min_{S'\subseteq S^\dag\cup S}\{f(S')-x(S')+d((S^\dag\cup S)\setminus S')\} \nonumber\\ 
	&= \min_{S'\subseteq S^\dag\cup S}\{f(S')-\tilde{x}(S')+\tilde{d}((S^\dag\cup S)\setminus S')\}-\delta(S^\dag\cup S) 
	=f_{\tilde{x},\tilde{d}}(S^\dag\cup S)-\delta(S^\dag\cup S).
	\end{align}
	Note that the first inequality holds by $S^\dag\cup (S\setminus S^*)=S^\dag\cup S$ (from $S^{\dag} \supset S^*$). 
	The second equality follows from that if buyer $i\in S^\dag\cup S$ clinches $\delta_i$ unit of goods in line 3 of Algorithm 2, 
	this increases $\tilde{x}_i$ by $\delta_i$ and decreases $\tilde{d}_i$ by $\delta_i$.
	Also, the third equality is by (\ref{hypothesis}). 
 
	Combining the inequalities (\ref{X_aft}) and (\ref{X_bef}), we have
	\begin{eqnarray*}
	f_{\tilde{x},\tilde{d}}(S)-\delta(S)>f_{x,d}(S)\geq f_{\tilde{x},\tilde{d}}(S^\dag\cup S)-\delta(S^\dag\cup S).
	\end{eqnarray*}
	This means $\delta(S^\dag\setminus S)>f_{\tilde{x},\tilde{d}}(S^\dag\cup S)-f_{\tilde{x},\tilde{d}}(S)$. 
	By $S^\dag\setminus S\neq \emptyset$, this contradicts with Lemma \ref{prepare_inv}. Therefore,  (\ref{hypothesis}) holds just after Algorithm 2 has finished.
	\end{proof}
\end{document}